\documentclass[journal]{IEEEtran}

\usepackage{subcaption}
\usepackage{epstopdf}
\usepackage{amsmath}
\usepackage{todonotes}
\usepackage{booktabs}
\usepackage{multirow}
\usepackage{listings}
\usepackage[inline]{enumitem}
\usepackage{pgf,tikz,pgfplots,pgfplotstable,etoolbox,color,xspace}
\usetikzlibrary{circuits.ee.IEC}
\usetikzlibrary{arrows,automata,backgrounds,calc,chains,decorations,matrix,positioning,patterns,petri,shadows,shapes,fit}
\usetikzlibrary{decorations.pathmorphing,decorations.pathreplacing,decorations.markings}
\usetikzlibrary{backgrounds}	% drawing the background after the foreground
\usetikzlibrary{pgfplots.groupplots}
\pgfplotsset{compat=1.7}
\usepackage{fixltx2e}
\usepgfplotslibrary{statistics}
\usepackage{adjustbox} %For two column figures
\usepackage{datetime}
\usepackage{siunitx}
\renewcommand{\dateseparator}{-}
\newcommand{\todayiso}{\the\year \dateseparator \twodigit\month \dateseparator \twodigit\day ~~ \currenttime}
\usepackage{fancyhdr}

\usepackage{algorithm}% http://ctan.org/pkg/algorithms
\usepackage{algpseudocode}% http://ctan.org/pkg/algorithmicx
\usepackage{amssymb}
\usepackage{graphicx}
\usepackage{multirow}
\usepackage{url}
\usepackage{slashbox}
\usepackage{tikz,ifthen,etoolbox,color,pgfplots} 
\usepackage{gensymb}
\usepackage{booktabs}
\usepackage{enumitem}
\usepackage{mathtools}
\usetikzlibrary{arrows,automata,backgrounds,calc,chains,circuits.ee.IEC,decorations,decorations.pathmorphing,decorations.pathreplacing,decorations.markings,matrix,positioning,patterns,petri,pgfplots.groupplots,shadows,shapes,fit}

\usetikzlibrary{arrows,automata,backgrounds,calc,chains,circuits.ee.IEC,decorations,decorations.pathmorphing,decorations.pathreplacing,decorations.markings,matrix,positioning,patterns,petri,pgfplots.groupplots,shadows,shapes,fit}

\usepackage{circuitikz}
\usepackage{todonotes}
\usetikzlibrary{circuits.ee.IEC}
\usetikzlibrary{arrows,automata,backgrounds,calc,chains,decorations,matrix,positioning,patterns,petri,shadows,shapes,fit,spy}
\usetikzlibrary{shadows.blur}

\usetikzlibrary{decorations.pathmorphing,decorations.pathreplacing,decorations.markings}
\usetikzlibrary{backgrounds}  % drawing the background after the foreground
\begin{document}
\raggedbottom

\title{DeCoRIC: Decentralized Connected Resilient IoT Clustering \\
\thanks{This work was financially supported in part by the Singapore National Research Foundation under its Campus for Research Excellence And Technological Enterprise (CREATE) programme. With the support of the Technische Universit{\"a}t  {\"M}unchen - Institute  for  Advanced Study, funded by the German Excellence Initiative and the European Union Seventh Framework Programme under grant agreement n\degree~291763}}

% These commands are optional
%\acmBooktitle{Transactions of the ACM Woodstock conference}
%\editor{Jennifer B. Sartor}
%\editor{Theo D'Hondt}
%\editor{Wolfgang De Meuter}

%\titlenote[none]{}

%\author{\IEEEauthorblockN{Nitin Shivaraman} \\
%\textit{TUMCREATE, Singapore}\\
%nitin.shivaraman@tum-create.edu.sg \\
%\and \IEEEauthorblockN{Shreejith Shanker} \\
%%\IEEEauthorblockA{\textit{dept. name of organization (of Aff.)} \\
%\textit{Trinity College Dublin, Ireland}\\
%shankers@tcd.ie \\
%\and \IEEEauthorblockN{Arvind Easwaran} \\
%%\IEEEauthorblockA{\textit{dept. name of organization (of Aff.)} \\
%\textit{Nanyang Technological University, Singapore}\\
%arvinde@ntu.edu.sg \\
%\and \IEEEauthorblockN{Sebastian Steinhorst} \\
%%\IEEEauthorblockA{\textit{dept. name of organization (of Aff.)} \\
%\textit{Technical University of Munich, Germany}\\
%sebastian.steinhorst@tum.de
%}

\author{
	\IEEEauthorblockN{{Nitin Shivaraman$^{1}$}, {Saravanan Ramanathan$^{1}$}, {Shreejith Shanker$^{2}$}, {Arvind Easwaran$^{3}$}, {Sebastian Steinhorst$^{4}$}}\\	
	\IEEEauthorblockA{$^{1}$TUMCREATE Limited, Singapore,
		$^{2}$Trinity College Dublin, Ireland,
		$^{3}$Nanyang Technological University, Singapore,
		$^{4}$Technical University of Munich, Germany\\
	Email: \normalsize {$^{1}$\{nitin.shivaraman, saravanan.ramanathan\}@tum-create.edu.sg, $^{2}$shankers@tcd.ie, $^{3}$arvinde@ntu.edu.sg\, $^{4}$sebastian.steinhorst@tum.de}}
	\vspace{-0.9cm}
}% <- end of author section

%\renewcommand{\shortauthors}{author et al.}
\maketitle

%\IEEEtitleabstractindextext{%
\begin{abstract}
%\begin{abstract}
%Decentralized architecture is gaining traction as the network architecture of choice in many critical Internet of Things (IoT) applications such as smart energy. Key requirements in such an IoT network are resilience to tolerate faults, decentralization for reliability and connectivity for peer-to-peer routing among all nodes with minimal power overhead. Although existing clustering approaches independently solve these issues, they are unsuitable for providing resilience and connectivity in an ad-hoc network with low power consumption.
%rendering clustering approaches from traditional wireless networks unsuitable. 
%We present Decentralized Connected Resilient IoT Clustering (DeCoRIC), a clustering scheme for IoT networks that is self-organizing and resilient to network changes while aiming for network connectivity. Using experimental evaluations implemented on the Contiki simulator, we show that our clustering scheme adapts itself to node faults in a time-bound manner. Our experiments show that DeCoRIC strives to achieve 100\% connectivity among all nodes, which is paramount for safety-critical systems. Our results also show that DeCoRIC improves power efficiency of nodes in the system by up to 110\% compared to the technique BEEM and 70\% over LEACH.
%The improved power efficiency also translates to longer lifetime before first node death with up to 109\% longer than BEEM and 42\% longer than LEACH.
%With the improved power efficiency, we are able to achieve up to 53\% longer time until the first node's death in the network.
Maintaining peer-to-peer connectivity with low energy overhead is a key requirement for several emerging Internet of Things (IoT) applications. It is also desirable to develop such connectivity solutions for non-static network topologies, so that resilience to device failures can be fully realized. Decentralized clustering has emerged as a promising technique to address this critical challenge. Clustering of nodes around cluster heads (CHs) provides an energy-efficient two-tier framework for peer-to-peer communication. At the same time, decentralization ensures that the framework can quickly adapt to a dynamically changing network topology. Although some decentralized clustering solutions have been proposed in the literature, they either lack guarantees on connectivity or incur significant energy overhead to maintain the clusters. In this paper, we present Decentralized Connected Resilient IoT Clustering (DeCoRIC), an energy-efficient clustering scheme that is self-organizing and resilient to network changes while guaranteeing connectivity. Using experiments implemented on the Contiki simulator, we show that our clustering scheme adapts itself to node faults in a time-bound manner. Our experiments show that DeCoRIC achieves 100\% connectivity among all nodes while improving the power efficiency of nodes in the system compared to the state-of-the-art techniques BEEM and LEACH by up to 110\% and 70\%, respectively. The improved power efficiency also translates to longer lifetime before first node death with a best-case of 109\% longer than BEEM and 42\% longer than LEACH.
\end{abstract}
%}

%%
%% The code below should be generated by the tool at
%% http://dl.acm.org/ccs.cfm
%\begin{CCSXML}
%<ccs2012>
%  <concept_id>10010520.10010553.10010562</concept_id>
%  <concept_desc>Computer systems organization~Embedded systems</concept_desc>
%  <concept_significance>500</concept_significance>
%  <concept_id>10010520.10010575.10010755</concept_id>
%  <concept_desc>Computer systems organization~Redundancy</concept_desc>
%  <concept_significance>300</concept_significance>
%  <concept_id>10010520.10010553.10010554</concept_id>
%  <concept_desc>Computer systems organization~Robotics</concept_desc>
%  <concept_significance>100</concept_significance>
%  <concept_id>10003033.10003083.10003095</concept_id>
%  <concept_desc>Networks~Network reliability</concept_desc>
%  <concept_significance>100</concept_significance>
%</ccs2012>
%\end{CCSXML}
%
%\ccsdesc[500]{Computer systems organization~Embedded systems}
%\keywords{ACM proceedings, \LaTeX, text tagging}

%\maketitle

%\let\thefootnote\relax\footnote{\small This work was supported by the Singapore National Research Foundation under its Campus for Research Excellence And Technological Enterprise (CREATE) programme.}

\section{Introduction}
\label{sec:intro}
 
%Such systems make informed decisions based on data from these connected devices. 
%The individual nodes in an IoT system use commercial processing platforms and interact over standard communication protocols.
The Internet of Things (IoT) enables millions of nodes (devices) to exchange information to form intelligent connected systems.
However, IoT networks exhibit some unique traits compared to `traditional' distributed systems. 
The nodes of an IoT network are primarily low-cost and resource-constrained, which may join or leave the network in an ad-hoc manner and communicate over a wireless interface.
Based on the information exchanged among neighboring nodes, each node may take independent decisions enabling decentralized operation.

Information exchange over a wireless medium necessitates the design of energy-efficient communication strategies for these nodes to extend their lifetime. Furthermore, it is a challenge to maintain end-to-end connectivity among the nodes of the network in the presence of changes in the communication links due to the ad-hoc nature of the network.
%While tackling energy consumption is a challenge in these systems, the network has to be flexible to adapt to the volatility in the communication links such that the network connectivity is maintained.

%IoT networks are inherently decentralized in nature as the network connectivity is not controlled by a central and often involve nodes with different processing capabilities. 
%The network schemes must ensure efficient and deterministic communication as well as ensure prolonged and sustainable operation under different and often challenging conditions. 
%Information flow in critical systems (presence of safety issue, security flaw or energy disruption), necessitates maintenance of communication links so that the data can be relayed from one part of the network to other parts.
%Maintaining this connectivity among all nodes of a network with volatility requires resilience to changes with communication links to re-adapt the connectivity.
%We explore 
%Hence, IoT networks not only need to be energy efficient but be resilient to changes in topology.
\begin{figure}[t]
\begin{center}
\includegraphics [width=0.99\columnwidth]{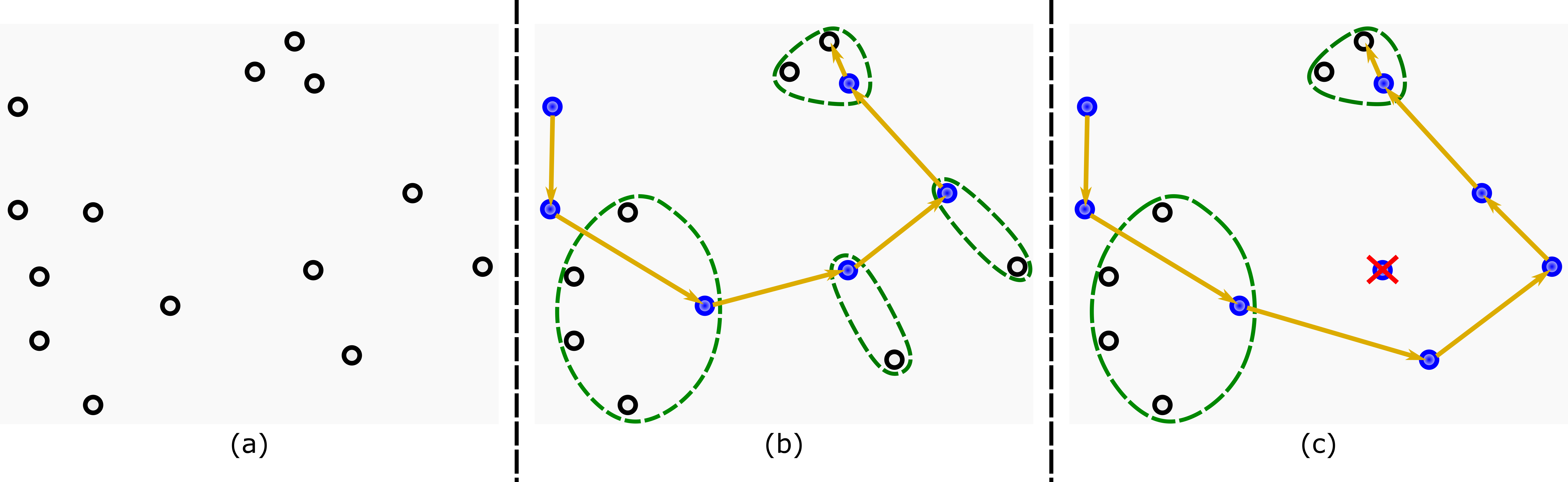}
%\vspace{1mm}
\caption{Clustering operation: starting with any topology (a), the nodes align themselves into clusters with an elected CH (b). With DeCoRIC, the clustering dynamically adapts to changes in topology to ensure connectivity among all nodes (c).}
\label{fig.concept}
\end{center}
\vspace{-5mm}
\end{figure}
 
Clustering has been shown as the most effective technique to improve energy efficiency and scalability in networked systems~\cite{926982}.
Nodes, as illustrated in Figure~\ref{fig.concept}\,(a), are grouped into \emph{clusters} based on common node properties such as residual energy, location or degree (number of communication links of the node).
Cluster sizes can be equal or unequal depending on the chosen property.
Nodes marked in blue are elected representative nodes called cluster heads (CHs), each of which acts as the data aggregator and nodal point for multi-hop communication, as shown in Figure~\ref{fig.concept}\,(b), allowing regular (non-CH) nodes to operate in low-power mode more often to conserve energy.
%\changes{Research has shown that unequal clusters tend to have better energy efficiency and can handle a large volume of data transmissions~\cite{1420160}.}

Most clustering techniques aim at improving the lifetime of the network, but often result in disassociated clusters and nodes operating independently without being able to establish any communication with neighbors, leading to loss of data and connectivity.
Also, static topologies lack the flexibility to deal with the ad-hoc nature of an IoT network as well as with node failures during operation. 
%Such networks are unable to accommodate new nodes or deal with failure of nodes at run-time. 
%In this paper, we propose a Decentralized Resilient IoT clustering (DeRIC) technique that can react to changes in the network/node conditions at run-time to ensure resilient network with maximal connectivity.
%Hence, in addition to the capabilities of the existing clustering techniques, we believe that the following properties are essential in any IoT clustering technique: 
We believe that the following properties are essential in any IoT clustering technique to complement the existing capabilities: 
\begin{enumerate}
\item \textbf{Connectivity} -- the clustering technique must ensure that the nodes are clustered such that there is a path between any two nodes in the network whenever possible; this property ensures reliable routing of information between any two nodes in the network.
\item \textbf{Decentralization} -- each node must make independent decisions without a central entity; this ensures there is no single-point of failure.
\item \textbf{Resilience} -- the network must adapt to node faults or network changes at run-time by detecting and reorganizing in a time-bound manner to ensure \emph{connectivity}.
\end{enumerate}

In this paper, we propose Decentralized Connected Resilient IoT Clustering (DeCoRIC), a clustering scheme that can group nodes into connected clusters and adapt to network changes at runtime without relying on a central node or prior information (topology, position, etc.).
Each node takes decisions independently and collectively manages the clustering process to achieve the above goals.
Based on the information gathered from their neighbors, nodes make decisions and react to topology changes by altering their state, as seen in Figure~\ref{fig.concept}\,(c), to ensure connectivity, while minimizing energy overheads. % dynamically at each node based on information it has about its neighbors and can thus

LEACH~\cite{926982} and BEEM~\cite{6878886} are chosen as the representative schemes for comparison. LEACH is the de-facto benchmark of clustering algorithms, while BEEM is a recent extension of another benchmark, Hybrid Energy-Efficient Distributed clustering (HEED)~\cite{Younis:2004}, aimed at higher connectivity. We evaluate DeCoRIC using multiple random network topologies to show that the above properties are achieved, while also enabling up to 70\% and 110\% improvement in power efficiency over BEEM and LEACH, respectively.

This paper presents the following contributions:
\begin{enumerate*}[label=(\roman*), itemjoin={{. }}, itemjoin*={{. }}]
	\item We propose DeCoRIC, our scheme for Decentralized Connected Resilient IoT Clustering (Section~\ref{sec.design})
	\item We have ported the state-of-the-art techniques LEACH and BEEM into the Contiki simulator to emulate a realistic communication environment (Section~\ref{sec.experiments}) for comparison with DeCoRIC and made the implementations open-source
	\item We show through results (Section~\ref{sec.experiments}) that DeCoRIC converges to a resilient fully connected network with bounded latency.
%DeCoRIC achieves up to 104\% higher power efficiency and 53\% longer lifetime compared to LEACH \changes{while achieving up to x\% better power efficiency and y\% longer network lifetime in comparison to BEEM.}% with significantly better energy efficiency over existing decentralized schemes.
\end{enumerate*}
\section{Related Work}
\label{sec.relatedwork}
Radio communication is a key component that largely influences the energy consumption in IoT nodes. 
Clustering techniques aim at reducing this energy consumption by partitioning the network into clusters.
Each cluster has an active cluster head (CH) as a representative node elected by either by a central entity or all the nodes in the cluster.
Clustering enables the regular (non-CH) nodes to reduce the frequency of transmission and operate in low-power mode, reducing the overall power consumed by the system. 
Cluster head oversees the multi-hop communication and performs data fusion resulting in minimal communication for the non-CH nodes.
Clustering techniques can be classified into centralized and decentralized methodologies, based on whether the clustering decision and CH election is performed by a central entity or independently by nodes of the network.

\begin{table*}[t]
	\begin{center}
	\caption{Comparison of notable works in Literature.}
	\label{tab.litcomp}
	%\begin{tabular}{llllllllllll}
	\begin{tabular}{|m{2.5cm}|m{1.8cm}|m{1.8cm}|m{1.8cm}|m{1.8cm}|m{1.8cm}|m{1.8cm}|m{1.8cm}|}
		\hline
		\vspace{1mm}
		Property/ protocol         & LEACH~\cite{926982}           & PEGASIS~\cite{1036066}                      & HEED~\cite{Younis:2004}           & PEACH~\cite{YI20072842}                        & EEUC~\cite{1542849}                                          & BEEM~\cite{6878886}               &  DeCoRIC       \\
		\hline
		\vspace{1mm}
		Location/ topology data    & No              & Yes                          & No             & Yes                          & Yes                                           & Yes                & No            \\
		\vspace{1mm}
		Centralized/ Decentralized & Decentralized   & Centralized                  & Decentralized  & Decentralized                & Decentralized                                 & Decentralized      &  Decentralized \\
		\vspace{1mm}
		Complexity                & Low             & High O(N\textasciicircum{}2) & Low             & High O(N\textasciicircum{}2) & Low                                           & Low                & Low           \\
		\vspace{1mm}
		Clustering mechanism      & Residual energy & Location                     & Residual energy & Proximity (overhearing)      & Residual energy and Base station proximity     					& Residual energy and Degree & Degree        \\
		\vspace{1mm}
		Communication channel     & TDMA            & CDMA                         & TDMA            & TDMA                         & TDMA                                          & TDMA        &  CSMA          \\
		\vspace{1mm}
		Connected clusters        & No              & No                           & Yes             & No                           & Yes                                           & Yes                &  Yes           \\
		\vspace{1mm}
		Resilience to failures    & No              & No                           & No              & No                           & No                                            & Yes                &  Yes           \\      
		\hline     
	\end{tabular}
\end{center}
\vspace{-5mm}
\end{table*}

\paragraph*{Centralized Clustering}Centralized techniques rely on a central entity that has global knowledge of the network, and manages the CH election and clustering process. 
The clustering operation can be based on the degree of a node in the network~\cite{Hartuv:2000}, residual energy of nodes~\cite{6840926} or other parameters. 
The degree of a node is defined as the number of neighbors within the radio range.
The clustering problem was formulated as a linear programming problem in~\cite{Ari:2016}, representing a trade-off between energy consumption and the quality of communication. 
A centralized version of a popular decentralized algorithm, LEACH (described below), was developed in~\cite{1045297}, where control decisions are managed by a central entity, making more efficient CH selection than LEACH.
Further improvement was made in~\cite{1036066}, where nodes are organized into a chain based on proximity to evenly distribute the transmission energy.
While there is no deterministic polynomial algorithm that can partition a network topology into clusters~\cite{avril2014clustering}, meta-heuristic algorithms like particle swarm optimization and artificial bee colony have been successfully applied in the clustering of wireless networks~\cite{elhabyan2014pso,karaboga2012cluster}.
%{Bee-Sensor-C~\cite{beesensorc} implement a localized event-based clustering by grouping nodes around an event (change in sensing parameter); sensors that are not in proximity to the event are left independent without clusters. Thus, this mechanism of clustering is not energy efficient.}
The requirement of a central entity (often the base station) in centralized systems results in higher clustering latency and scalability issues, since every decision has to be relayed from the central entity. %, which are not ideal in case of critical systems with dynamic networks. 
%This inhibits resilience and connectivity without a central entity. 
%Decentralization mitigates some of these issues.
%However, optimized centralized techniques offer better energy efficiency over distributed clustering schemes.
The centralized approach also results in a single point of failure at the central node, inhibiting effective ways to enable resilience and connectivity.
Distributing the clustering operation among nodes aims to mitigate some of these issues, albeit centralized techniques generally offer superior energy efficiency over decentralized strategies.

\paragraph*{Decentralized Clustering}HEED was one of the earliest decentralized techniques and uses a combination of node degree and residual energy as the metric for clustering~\cite{Younis:2004}.
%PASCAL~\cite{Mirza:2009} improved the power efficiency of HEED through multi-level sectoring.
BEEM~\cite{6878886} is the most recent extension of HEED that includes node degree in the CH election conditions to improve connectivity by letting nodes in denser areas expend higher energy.

Low-Energy Adaptive Clustering Hierarchy (LEACH)~\cite{926982} is the most popular decentralized technique, which used probabilistic election of a CH and its rotation within a cluster to ensure uniform energy distribution.
%The Clustering metric was based on the residual energy of a node and its probability of becoming a CH, ensuring uniform energy distribution among the cluster nodes. 
Enhancements to the LEACH protocol that enable power optimization through two-level adaptive clustering~\cite{loscri2005two}, and multi-level hierarchical clustering~\cite{1209194} have also been proposed.
More recent enhancements to the protocol added residual energy~\cite{8633905} and multi-hop communication~\cite{ALSODAIRI20181} in the CH election process to achieve minor improvements in energy and throughput, respectively.
%\cite{8633905} creates a 2-stage process with first stage reusing LEACH and second stage adding residual energy information to update the CH.
LEACH and HEED are used as benchmarks in clustering by the community~\cite{7976279}. % and hence we compare our proposed algorithm with LEACH and BEEM (extension of HEED).
As BEEM extends HEED ensuring connectivity, LEACH and BEEM are chosen as representative techniques in our paper for comparison with DeCoRIC.
%Ring-Structured Energy-Efficient Cluster Architecture (RECA)~\cite{reca} adds a dynamic re-clustering scheme that alters the clusters if the energy falls below a preset threshold. % the energy of the cluster falls below a threshold.

Other notable works include overlapping clusters~\cite{4796192, ace} where nodes belong to multiple clusters simultaneously to ensure connectivity among the clusters. 
Unequal clusters~\cite{1542849} were used to reduce the impact of high activity for nodes close to the base station.
The work in~\cite{YI20072842} form multi-level clustering using overhearing characteristics of the wireless medium to form clusters adaptively.
%Algorithm for Cluster Establishment (ACE)~\cite{ace} also employs an overlapping strategy for connectivity with CH election based on degree of node and supports for ad-hoc network architecture creating high overlap among the clusters. 
Event-based clustering schemes such as Bee-Sensor-C~\cite{beesensorc} perform a local clustering around an event (such as sensor value change) but suffer from poor energy efficiency without any clustering for non-event nodes.
%supports ad-hoc network structure, with CHs determined using the degree of a node 
%Moreover, some techniques like~\cite{reca} rely on pre-elected CHs to achieve energy efficiency.
%Few methods use multi-hop from CH to the leaf nodes causing higher node activity.
Methods such as~\cite{1209194} use multi-hop within the cluster, causing high node activity and energy consumption, while also presenting challenges in reliable message delivery and clustering convergence when the network scales.
%Convergence to a stable operating mode cannot be achieved in a bounded time.
%As the number of nodes scale, there is an impact on reliable message delivery and power consumption, without converging to a stable operating mode in bounded time. 
Techniques that employ overlapping for connectivity~\cite{4796192,ace} are susceptible to hidden node collision faults, affecting reliable message exchange.
However, most of the existing decentralized schemes use a fixed network topology and cannot cater to dynamic ad-hoc networks of the IoT.
Further, most techniques do not consider connectivity across all nodes and often result in isolated clusters, albeit the nodes are within the radio range of each other. 
A summary of some works and their properties is shown in Table~\ref{tab.litcomp}.

Additionally, there is a body on literature which look into clustering in a graph theoretic perspective~\cite{JALLU2017159,7917339,7524504}.
The network is mapped as a unit disk graph to find the minimum connected dominating set (MCDS) for various network topologies.
However, the literature in this direction is unrelated to the presented algorithm in this paper as they do not consider any radio model in the network, leading to a theoretical solution that may not be practically viable.

To the best of our knowledge and as reported in the literature~\cite{7976279}, there is no existing clustering method that combines the three properties of decentralization, connectivity and resilience.

%Overlapping nodes that do not belong to an exclusive cluster (common among multiple clusters) suffer from hidden node collision issue.
%DeCoRIC provides a resilient implementation for catering to reliable message delivery as converges within a bounded time. 
%With independent nodes instead of overlapping nodes, the messages that are not meant for inter-cluster communication can be ignored mitigating the hidden node collision problems as described in Section~\ref{sec.design}.
%Finally, most algorithms do not ensure connectivity to all nodes and often end up with isolated clusters though nodes existing within these clusters are within radio range of each other (i.e., \textbf{connectivity} property).

%DeCoRIC aims to address these challenges while offering comparable energy performance to the centralized clustering schemes.

\section{DeCoRIC strategy}
\label{sec.design}
In this section, we will introduce the network assumptions as well as the detailed clustering phases of DeCoRIC.
\subsection{IoT Network Assumptions}
%The IoT network can be modeled as a connected graph defined by an ordered pair $\langle V ,E \rangle$, with $V$ denoting the set of vertices and $E\,\subseteq\,V$\,x\,$V$ denoting the set of undirected edges between the vertices. 
%The degree of a vertex $i$, represented as $D_i$, is the number of neighbors of $i$ in the graph. It is equivalent to the vertices that are within its radio range.
%%For any two vertices $u, v \in V$, there is an edge $e_{uv} \in E$ such that $e_{uv} = e_{vu} \forall e_{uv} \in E$ representing a bidirectional link forms an undirected graph. %\ToNitin{What does this mean?}
%The vertices of the undirected graph correspond to IoT nodes of a network, while the edges are the communication links among the nodes.
%\emph{Bridge nodes} act as connectors between different parts of a network whose edges are bridges of the graph. They are the nodes which have at least one bridge edge without which the network becomes disconnected. These nodes are essential to warrant connectivity of the network. 
%%Without whose nodes and its associated edges, the graph would results in a disconnected network.
%%Similar to critical vertices of a graph, there are nodes which act as connectors between two parts of the network are called bridge nodes. 
%A \emph{Bridge node} failure would result in a (partially) disconnected network.
We make the following assumptions about the IoT network:
\begin{enumerate}
	\item The network uses wireless communication among the nodes based on the IEEE 802.15.4 standard~\cite{1700009}.
	\item Carrier-sense multiple access with collision avoidance (CSMA/CA) is employed at the MAC layer to mitigate congestion for broadcast messages. % in the network.
	\item Each node operates independently without any central entity or apriori information about the topology (i.e., fully decentralized network).
	\item Nodes are equipped with a processor, clock, memory and a unique identification (ID) used for book-keeping.
	\item The network is ad-hoc where nodes can join or leave the network at runtime (e.g. node failures).
	\item All nodes transmit and receive on the same channel with the same signal strength asynchronously.
	\item The network uses a hard fault failure model, i.e., a faulty node ceases to transmit information on the network. %lie about their information.
\end{enumerate}

\subsection{Clustering Strategy}\label{sec.strategy}
\begin{figure}[tb]
\begin{center}
	\includegraphics[scale=0.95,width=0.99\columnwidth]{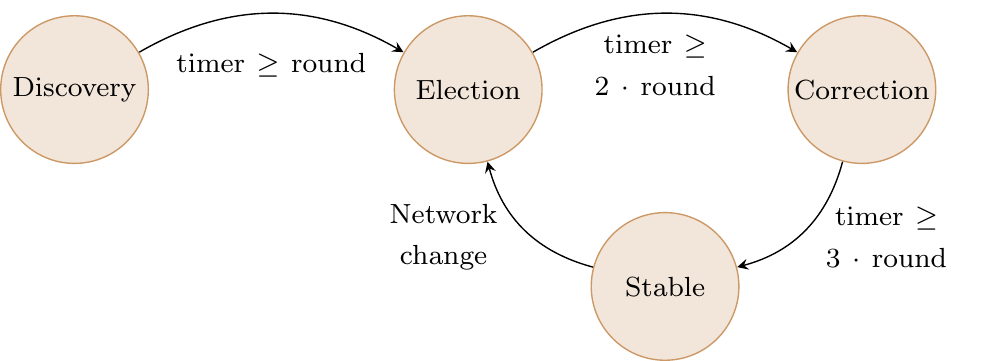}
	\caption{DeCoRIC phases and transition conditions.}
	\label{fig.statemachine}
	\end{center}
\end{figure}

% short description of the phases of clustering
DeCoRIC operates independently at each node in four phases defined by a Finite State Machine (FSM) as shown in Figure~\ref{fig.statemachine} with each phase lasting for a pre-configured period (called \emph{round}, discussed below). %Each phase represents the operating condition of a node in relation to its network, . 
%The pseudo-code of the strategy is given in Algorithms under the corresponding state and the state machine is given by Figure~\ref{fig.statemachine}.
%Each node runs this algorithm independently.
%\toshreejith{review the overview below}
The nodes power up in the \emph{Discovery} phase, where each node waits  to receive messages from its neighboring nodes and evaluate its environment. 
%\toshreejith{Is the below statement grammatically and logically correct? I replaced "evaluate" as it was used multiple times.}
The transition occurs at the end of the period to the \emph{Election} phase, where the node that has the highest degree declares itself as a CH followed by neighboring nodes associating themselves to nearby CHs forming clusters. %each node checks its status with respect to others. Following this evaluation, 
%The degree of a node is defined as the number of neighbors within radio range and thus can determine the potential for a node to act as a CH.
Progression to the \emph{Correction} phase after Election phase initiates evaluation of the connectivity property to identify isolated clusters/nodes. %occurs thereafter,where
Non-CH nodes within a cluster, which can enable connectivity between two CH nodes that are out of range, break out to form \emph{Bridge-CH} nodes. %and are currently associated with the
The system transitions into the \emph{Stable} phase at the end of the Correction phase, where the nodes periodically check the status of their neighbors by exchanging \emph{health} information.
If changes are detected (i.e., failures or new nodes in the system), the nodes go back to the Election phase and follow the path to re-establish a stable operation.
Once in the Stable phase, a Bridge-CH could upgrade itself to a full CH, if newly joining nodes affiliate themselves with the Bridge-CH, forming new clusters. %t} get added to the system which have proximity to the Bridge-CH instead to other CHs.

%Furthermore, if there are isolated clusters formed inhibiting the routing between clusters, the bridge nodes check some conditions and breakout to form independent CHs to facilitate data routing. Finally, in the stable phase, health messages are sent at different frequencies as long as there are no network changes.
%The different states of operation are described below.

\begin{figure}[tb]
	\begin{center}
		\includegraphics[width = 0.94\columnwidth]{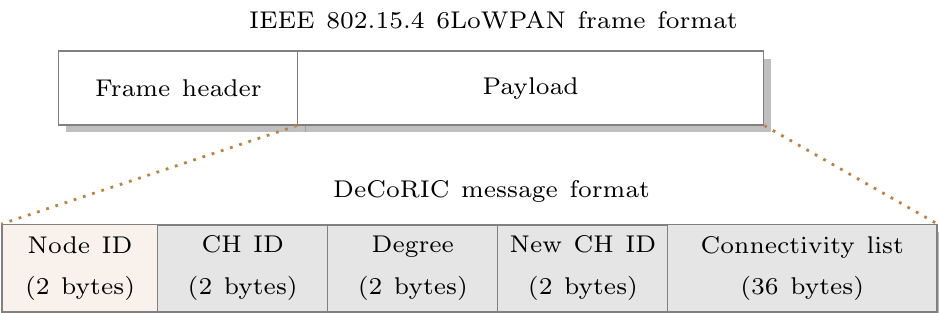}
		\caption{DeCoRIC message format.}
		\label{fig.msgformat}
	\end{center}
\end{figure}

% Describe the frame fields
To enable this operation, DeCoRIC uses broadcast messages as payload, shown in Figure~\ref{fig.msgformat}, that is encapsulated in a regular IEEE 802.15.4 frame.
The \emph{Node ID} field marks the ID of the transmitter and is always present in all messages. 
Only the Node ID field is valid in the Discovery phase as there is no information about the neighboring nodes. %; parts of the frame get filled based on the information each node has about its neighbors.
%In different phases, parts of the frame get filled based on the information each node has about its neighbors; he Discovery phase, only the Node ID field is valid as the node has no other information.
In the subsequent phases, the \emph{CH ID} and degree become known to each node, which feeds into the \emph{neighbor list} (i.e., a list of all neighbors a node has received direct messages from).
While the neighbor list contains all neighbors that are in range of the node, the actual connected nodes are maintained using a second list called the \emph{connectivity} list.
The connectivity list gets updated periodically, reflecting the activity of connected nodes. 
This two-level list structure allows DeCoRIC to eliminate false positives on the propagation of the activity of the nodes during the Stable phase (discussed in Section~\ref{phase4}).
%\changes{Since these neighbors enable connectivity among themselves, each node maintains a \emph{Connectivity} list which contains neighbors with active connections.
%A node in the connectivity list is updated every time a message from the corresponding node is received.}
The \emph{New CH ID} field is used only when a new node determines that it has to be the CH as it attained a higher degree than its current CH.
The operation details of each node as it transitions through the DeCoRIC phases are described below.

\begin{figure}[tb]
	\centering
	\includegraphics[width=1.0\columnwidth,trim=6cm 4.5cm 6cm 4.5cm,clip]{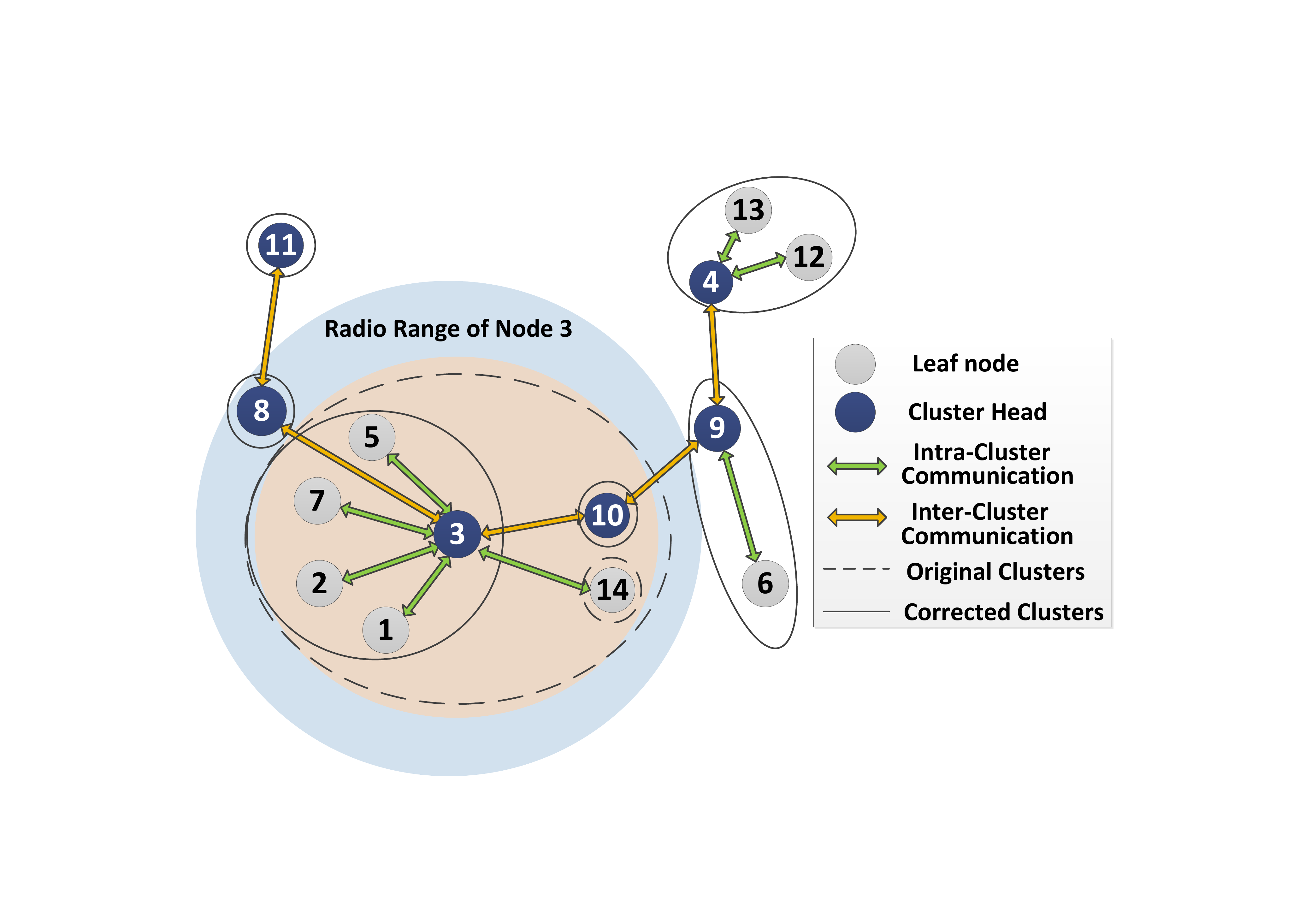}
	\caption{DeCoRIC on an example network (not drawn to scale).}
	\label{fig.cluster_eg}
\end{figure}

\subsubsection{Discovery phase}\label{phase1} 
% Basic setup and description of the phase.
The neighbor discovery phase enables each node to discover its neighboring nodes that it can communicate with and, hence, its own degree. 
The steps involved in the Discovery phase are listed in Algorithm~\ref{alg.cluster1}.
In this phase, each node sends a DeCoRIC \emph{ping} message with only the node ID and CH-ID fields filled with its own identifier (others left as zeros).
All nodes keep their radio active during this phase to receive messages from their neighbors.

% RSSI parameter update and clustering condition.
The RSSI threshold is a configurable parameter to ensure that communication links among the nodes can offer sufficient signal to noise ratio (SNR)~\cite{6785450} for reliable communication.
A receiving node updates its degree with each received message and updates the neighbor list. 
Nodes that meet a received signal strength indicator (RSSI) threshold are marked as a \emph{potential} neighbors that could belong to the same cluster.
Nodes that fail to meet the threshold do not belong to the same cluster and are marked as external neighbors.
%Each receiving node (within the signal range) saves the ID of the transmitter to the neighbor list. %whose ping message meets a configurable received signal strength indicator (RSSI) threshold as a \emph{potential} neighbor as shown in lines 3-9 of Algorithm~\ref{alg.cluster1}.
%Potential neighbors are nodes that can belong to the same cluster;

%\changes{The neighbor list of each node is populated with its neighboring nodes from which messages are received.}

% Example network
For the example network shown in Figure~\ref{fig.cluster_eg}, node 3 has a communication range of the blue shaded area while the RSSI threshold limits the cluster range to the orange shaded area. Node 3 receives messages from 1, 2, 5, 7, 8, 10 and 14 leading to a degree of 7. 
Node 8 is marked as an external neighbor while nodes 1, 2, 5, 7, 10 and 14 are marked as potential neighbors based on the configured RSSI threshold. %as these fall within the blue shaded region which defines the radio range of node 3.
%However, node 8 is beyond the RSSI threshold range of node 3 (marked by the dashed region), and thus is marked as an external neighbor as opposed to the other nodes.
Node 3 includes these 7 neighbors in its neighbor list at the end of this phase.
Similarly, node 9 marks 6 as a potential neighbor, while 4, 10, 12, 13, 14 become external neighbors to 9. 
Nodes 12 and 13 are marked potential neighbors by node 4, with node 9 as an external neighbor.
Node 11 is out of the RSSI threshold range of node 8 but within the radio range. Thus, node 8 marks 11 as an external neighbor.
%Nodes 1, 2, 5, 7, 8, 10 and 14 are in the radio range of node 3 as shown in Figure~\ref{fig.cluster_eg}. Since node 8 is beyond the RSSI threshold of node 3, it is marked as an external neighbor. Similarly, nodes 9 and 6 are in close proximity and nodes 4, 12 and 13 are in radio range of each other. Node 11 is out of radio range with any node of the network. Hence, it remains an independent node.
%The number of neighbors determines the degree of each node, and hence its potential to act as a cluster head.  
%In the example case, node 3 has a degree of 7 which includes both its potential as well as external neighbors.
%Similarly, node 9 has a degree of 6 with node 6 as its potential neighbor and nodes 4, 10, 12, 13, 14 as external neighbors.

% Why round is needed and round equation
Since the nodes in the network communicate asynchronously, multiple nodes will attempt to transmit during any given time, resulting in collisions.
Although the use of CSMA-CA avoids collisions, it is important that a node is able to complete the transmission without failures or indefinite wait times due to back-off.
To ensure that each node can transmit at least one DeCoRIC message in the Discovery phase, the transmission time window is computed based on the IEEE 802.15.4 standard~\cite{alvi:hal-01434866} including the worst-case back-off. 
This value is aggregated over the maximum number of nodes supported by the network to form a time window referred to as \emph{round} in DeCoRIC. %, calculated using the Equation~\ref{roundEQ}.
The nodes stay in the Discovery phase for one round, which ensures that all nodes have successfully transmitted at least one DeCoRIC \emph{ping} message.
It is important to note that all nodes are asynchronous. 
%All nodes keep the receiver active in this phase and receive all valid transmissions from nodes within their radio range.  
The time duration of one round is calculated as: 
%\begin{equation} \label{roundEQ}
%\begin{split}

\begin{equation}
round = N \cdot (\sum_{i=0}^{maxR} \tau_{\mathrm{bo,i}} + \tau_{\mathrm{fr}} + \tau_{\mathrm{ifs}})\
\label{eq:round} 
\end{equation}
%\text{where} & \\
%& N = \text{Maximum number of nodes} \\
%& t_{trans} = \text{Transmission time for a packet}
%\end{split}
%\end{equation}
%\label{eq:backoff}
\[\text{where,} ~\tau_{\mathrm{bo,i}} = (2^{maxBE(i)}-1) \cdot \tau_{\mathrm{symb}} + 2 \cdot \tau_{\mathrm{cca}}, \]$N$ is the maximum number of nodes in the network, $maxR$ is the maximum number of retries, $\tau_{\mathrm{bo,i}}$ is the worst-case back-off delay at the \emph{i}\textsuperscript{th} retransmission, $\tau_{\mathrm{fr}}$ is the frame transmission time, $\tau_{\mathrm{ifs}}$ is the minimum inter-frame period, $maxBE$ is the maximum back-off parameter at the \emph{i}\textsuperscript{th} retransmission, $\tau_{\mathrm{symb}}$ is the back-off symbol period and $\tau_{\mathrm{cca}}$ is the clear channel assessment time. 
The parameters $\tau_{\mathrm{ifs}}$, $\tau_{\mathrm{symb}}$ and $\tau_{\mathrm{cca}}$ are derived from the network standard (IEEE 802.15.4), while $maxBE$, $N$ and $\tau_{\mathrm{fr}}$  are configured with the same value at each node (as network parameters).
%A round is the fundamental time window in DeCoRIC that is configured to be wide enough to accommodate at least one transmission based on maximum number of nodes in the network.
%\toshreejith{Description about collisions of nodes and no assumption on nodes being static}

\begin{algorithm}[t!]
	\scalebox{0.90}{
	\begin{minipage}{0.98\columnwidth}
	\caption{Neighbor Discovery phase}
	\label{alg.cluster1}
	\begin{algorithmic}[1]
		\State LIST: {Neighbor, Conn} = FALSE
		\State Degree = 0, CH.ID = node.ID
		\State broadcast(Msg)
		\If{rcv()}
		\State Msg = rcv().data
		\State Neighbor[Msg.ID], Conn[Msg.ID] = TRUE
		\State Degree = Degree + 1
		\If{rcv().RSSI $<$ RSSI\_threshold}
			\State Neighbor[Msg.ID].ext = TRUE
		\EndIf
		\EndIf
	\end{algorithmic}
	\end{minipage}}
\end{algorithm}
 
%
%A special case is when a new node tries to integrate into an existing cluster (i.e., it observes \emph{health} messages instead of ping messages), in which case it waits for a minimum period of one \emph{cycle}.
%\tonitin{One round or one cycle?}
%At run-time a node addition can be by comparing the ID of the received message with its neighbor list irrespective of the state.

% The transmission strength and threshold is quantified using Received Signal Strength Indicator (RSSI). 
% Nodes having RSSI within the threshold become neighbors and are potential members of same cluster. 
% Meanwhile, nodes which are out of threshold could be members of different clusters or operate as independent clusters to ensure network connectivity. 
% Such nodes are denoted as external neighbors. 
% Cluster Heads communicate with other external CH nodes to transmit messages across the network to different clusters. 
% The RSSI threshold is a configurable parameter. 
% The threshold ensures that the received signals have a good quality. 
% Due to multiple nodes transmitting in parallel, there could be some nodes which may not be able to transmit. 
% Hence, empirically this phase is configured to be three rounds to make sure all the nodes have received at least one message from their respective neighbors. 
% Each round is defined to be transmission of a message from every nodes while each cycle is defined to be 5 rounds.

\subsubsection{Election phase}
\label{phase2} 
In this phase, nodes transmit a DeCoRIC message with an up-to-date degree field obtained from the previous phase. %to other nodes, in order to determine the cluster head. 
Each receiving node independently compares its own degree to the received degree to keep track of the node with the highest degree (potential CH).
Once each node has received at least one transmission from each neighbor (ensured by round configuration), it sets the node with the highest degree as its CH. %(similar scheme to~\cite{Wen2005,ace}). 
%Thus, each node in the system affiliates itself to a CH, and forms a cluster following the min-cut~\cite{Stoer:1997} strategy, with the cut defined by the degree of each node. 
If multiple nodes have the same highest degree, the node with the lower node ID is chosen as CH. %as seen in lines 2-6 of Algorithm~\ref{alg.cluster2}.
The configuration can be altered to choose a higher ID or support priority for specific node IDs. % (can be altered by setting a parameter).
The operation of this phase is described in Algorithm~\ref{alg.cluster2}.
%The elected CH oversees communication within its cluster and acts as the gateway for inter-cluster communication.
%Each node compares its degree to the received degree and associates itself with the sending neighbor if its degree is higher. 
%If a node's degree is highest than its neighbors' degree, it declares itself to be the Cluster head. Essentially, the cut in the graph is made based on the node degree. 
%The node with the highest degree is elected as CH; if multiple nodes have the same degree, the node with the lower ID is chosen as CH (can be configured by setting a parameter). 
%The election of CH is similar to~\cite{Wen2005, ace}. \toshreejith{Once elected, the CH is responsible to oversee communication within the cluster and also acts as the gateway for inter-cluster communication. $=>$ Do we include this statement as there is no routing?}
%This phase is also configured into one round to allow all nodes to transmit their messages. The radio continues to remain active during this phase.
Similar to the discovery phase, all nodes keep the radio active during this phase. %, and thus 
A message from a new node will be updated into the neighbor list and the connectivity list, while messages from existing nodes reinforce their active state in the connectivity list.

In the case of the example system in Figure~\ref{fig.cluster_eg}, node 3 becomes a CH with nodes 1, 2, 5, 7, 10 and 14 as its members at the end of this phase. 
Node 9 becomes a CH with node 6 as a member, node 4 forms the CH with nodes 12 and 13 as members, while nodes 8 and 11 become independent CHs. 
%This phase can also be completed in a minimum period of one round similar to previous state as the nodes keep their receiver active during this phase. 
%Node 3 becomes a CH with nodes 1, 2, 5, 7, 10 and 14 in its cluster. Node 9 declares itself as a CH with node 6 as a leaf node in its cluster. Node 4 forms the CH of the cluster comprising of node 12 and 13.
%Since the CH can be determined with one DeCoRIC message from each node in the system, the clusters can be formed within one round and the nodes transition into the next phase at the end of this period.

\begin{algorithm}[t!]
	\scalebox{0.90}{
	\begin{minipage}{0.98\columnwidth}
		\caption{Cluster Election phase}
		\label{alg.cluster2}
		\begin{algorithmic}[1]
		\State broadcast(Msg)
		\If {rcv()}		
		\State Msg = rcv().data
		\If{Msg.Degree $>$ Degree} \, CH.ID = Msg.ID
		\ElsIf{(Msg.Degree == Degree) \& (node.ID $>$ Msg.ID)} 
		\State	CH.ID = Msg.ID
		\EndIf
		\EndIf
		\end{algorithmic}
	\end{minipage}}
\end{algorithm} 

% Node degree obtained in the previous phase is sent to its neighbors by each node to determine the node with highest degree. 
% This is similar to min-cut~\cite{Stoer:1997} where the network is clustered such that each node associates itself with a node having maximum degree. 
% The node with the highest degree becomes the Cluster Head (CH) and all nodes within its communication range become cluster members. 
% If there are multiple nodes having the same degree, the node with a lower ID is chosen as CH by convention. 
% The configuration can be also be updated to select the node with higher ID in the event of contention for CH. 
% CH is responsible to oversee communication within a cluster and acting as a gateway to communicate with other clusters. 
% This phase is also configured to be three rounds similar to neighbor discovery to allow any node that has missed transmission to complete.

\subsubsection{Correction phase}\label{phase3} 
%Once the initial clusters are formed, each node verifies a set of conditions to ensure that no node within the radio range is isolated. 
Once the clusters are established, there exists the possibility of isolated clusters, i.e., the Cluster Heads are not within each others' radio range, but some common nodes of either cluster can connect the two clusters.
%Isolated clusters occur when nodes in different clusters are within each other's radio range, while their respective CH's are not.
To prevent isolated clusters, each non-CH node verifies the \textbf{connectivity} property based on the connectivity list and root-ID fields of the received messages. 
If any non-CH node satisfies the connectivity property, it breaks out from the affiliated cluster to form a Bridge-CH. % observes that it can connect 2 CH's which are not in each other's
This correction process is described in Algorithm~\ref{alg.cluster3}. 
If multiple nodes can enable connectivity between the same set of CHs, the rule for CH election is followed (see Section~\ref{phase2}).
Redundant bridge nodes continue to operate as non-CH nodes, reducing interference to the existing bridge nodes during inter-cluster communication and, thereby, minimizing their power consumption.
Once all the nodes verify the connectivity property, the clusters and CHs established are finalized and the network moves to a stable execution phase.

%This condition is shown in an example scenario in Figure.~\ref{fig.cluster_eg} with 13 nodes (marked 1 to 13). 
%As described in section~\ref{sec.relatedwork}, it is energy efficient for the bridge node to break out into a separate cluster head to facilitate communication between two cluster rather than being part of two clusters at the same time. 
%The shaded ring in the figure marks the radio range of node 3. 
Referring back to the example in Figure~\ref{fig.cluster_eg}, nodes 10 and 14 identify that they can enable direct connectivity between CH node 9 and their current CH node 3 based on information from the connectivity list. %could form a bridge node (and a CH) connecting the isolated clusters. %forming a fully connected network.
%following the first 2 phases, nodes 10 and 14 form a cluster with node 3 as the CH (shown by the dashed boundary), while node 6 forms a cluster with node 9 as CH at the end of Election phase. In this phase, 
As node 10 has the same degree as node 14, node 10 breaks out as the Bridge-CH because of its lower ID, while node 14 continues as cluster member. %(following rule for CH election described in sec.~\ref{phase3}). 
%The shaded area represents the radio range of node 3 and the dotted line represents the cluster before node 10 breaks out. Node 11 being out of radio range of other nodes, becomes an independent cluster.
%A Bridge-CH could migrate to a full CH if new nodes get added to the system which have higher proximity to the Bridge-CH than other CHs. % (i.e., bridge node transmissions have the highest RSSI value at the new node).
%If there are new nodes added to the network, they could have bridge nodes as their CH if the bridge node has better RSSI than the other CHs.
%
%The pseudo-code of DeCoRIC over these phases is shown in Algorithm~\ref{alg.cluster}. 
%Each node runs this algorithm independently and thereby, decisions made in the network is fully decentralized.

The correction phase ensures that non-CH nodes strictly remain in low-power mode without involvement in inter-cluster communication while connectivity among nodes is ensured by CH nodes.
Further, this phase minimizes energy overhead and latency in inter-cluster communication by enabling single-hop connection among CHs, while lowering congestion and error propagation at the bridge-CH interfaces (hidden node collision problem)~\cite{Tseng:2014}.
A node which breaks out from a cluster will not attempt to reintegrate into a cluster and remains an independent cluster head unless a network change invalidates the correction.
This ensures that the algorithm converges to an operative network condition at each node in a time-bound manner without frequent re-clustering.

\begin{algorithm}[t!]
	\scalebox{0.90}{
	\begin{minipage}{0.98\columnwidth}
		\caption{Cluster Correction phase}
		\label{alg.cluster3}
		\begin{algorithmic}[1]
		\State broadcast(Msg)
		\If {rcv()} 
		\State	Msg = rcv().data
		\If{(Msg.CHID == Msg.ID) AND (Msg.CHID != CH.ID)} 
			\If {Msg.Conn[CH.ID] == 0}
					\State CH.ID = node.ID \, /* Detach from cluster */
			\EndIf
		\EndIf
		\EndIf
		\end{algorithmic}
	\end{minipage}}
\end{algorithm}

%  two nodes are communicating eliminating power consumption by other nodes and frame data has lesser error induction with fewer nodes communicating. 
% Second condition ensures there are no isolated clusters as described earlier. 
% If either of the two conditions are not satisfied for a node, then the node breaks out of the cluster to become an independent cluster irrespective of the communication reach with its former CH. 
% The independent cluster acts as a bridge to connect two different clusters. 
% This operation can be shown in the example in Figure~\ref{fig:1} and Algorithm~\ref{alg:cluster}. 
% As seen in the example, node 10 forms a cluster with node 3 as CH. Since CH node 3 cannot access CH node 9 forming isolated clusters (shown by the dashed boundary).
% Node 10 satisfies the above conditions to break out of the cluster to become an independent cluster connecting the Clusters Heads node 3 and node 9. 

%\ToNitin{Otherwise, the nodes affected by phase~3 try to re-cluster iteratively, creating new clusters and/or CHs to reach a stable network.} % which will lead to a stable connected network.

\subsubsection{Stable phase}\label{phase4} 
In the Stable phase, the CH nodes broadcast a fully populated DeCoRIC frame as \emph{health} message every round.
Health message informs the other nodes that the sender node has not exhausted its energy.
They also serve as a re-clustering trigger if there is a change in the network topology.
All nodes activate radio duty cycling (RDC)~\cite{Dunkels:2004} which keeps the receiver active only for a fraction of time in a periodic manner (determined by parameter $RDC_{\mathrm{rate}}$) to reduce the power consumed by the radio.
Thus, a successful transmission may not guarantee reception at each node.
To address this, DeCoRIC defines a configurable period called \emph{cycle} as the minimum set of rounds that will ensure that a non-CH node receives at least one transmission from its CH. %another configurable
%\changes{RDC is employed in CSMA based networks to reduce the active time of the transceiver.
%Use of CSMA over TDMA is justified for scalability of networks as TDMA can only accommodate limited slots.
%With CSMA, the transmitted frame length is flexible along with the underlying protocol (WiFi, 6LoWPAN, etc.), whereas, while frame length impacts the slots in TDMA.
%A successful transmission in DeCoRIC is ensured through the selection of round as described in Equation~\ref{eq:round}.}
%The cycle duration is computed based on Equation~\ref{cycleEQ}.
%For a network of 100 nodes, the round and cycle parameters can be derived as 1.1\,seconds and 6\,rounds; any multiple of these values also being valid network configurations in DeCoRIC. 
Cycle duration is computed as: 
\begin{equation}
cycle = h \cdot \text{LCM}(txn_{\mathrm{freq}}/h, RDC_{\mathrm{rate}}/h)\
\label{eq:cycle}
\end{equation}
\[\text{where,} ~h = \text{GCD}(txn_{\mathrm{freq}}, RDC_{\mathrm{rate}}), \]

$txn_{\mathrm{freq}}$ is the round duration, $RDC_{\mathrm{rate}}$ is the RDC frequency in number of ON periods per second, GCD() and LCM() are the greatest common divisor and least common multiple functions, respectively. 
%\begin{equation} \label{cycleEQ}
%\begin{split}
%cycle & = h \cdot \text{LCM}(txn_{freq}/h, RDC\_rate/h), \\
%\text{where}\,h & =  \text{HCF}(txn_{freq}, RDC\_rate) \\
%%txn\_frequency & = round_{min}
%\end{split}
%\end{equation}
%and $txn_{freq}$ is the round configuration.\\

Non-CH nodes aggregate the received CH health messages over a cycle and acknowledge with a health message at the end of the cycle.
%The acknowledgment message includes the ID of the leaf node and its own connectivity list (i.e., the list of neighbors from which a \emph{direct} message was received). 
The per-cycle message from non-CH nodes not only reduces network traffic, but also conserves energy at these nodes.
The connectivity list aids in failure detection and gets updated upon receiving health messages.

%The health messages enable detection of failed nodes in the system and new nodes in the vicinity. % or nodes which have failed. 
%Each node maintains a \emph{fail} counter for every node in its neighbor list that increments at every round.
%$T_{\mathrm{fail}}$ is calculated from the equation for cycle with $txn_{\mathrm{freq}}$ set to 1 round for CH nodes and 1 cycle for non-CH nodes.
%The counter is reset when a health message is received successfully from the corresponding node.
%A neighbor is marked as faulty if the counter exceeds the time window of $T_{\mathrm{fail}}$. %node fails to receive at least one health message within
%This process is further explained in the next section.

\paragraph*{Failure Detection}

%each non-CH node (or leaf node) only transmits periodic health messages to ensure that it is still active. 
%This reduces the transmission power for leaf nodes as well as reduce the turn-on frequency for all the nodes as there are lesser messages on the medium.
%The CH nodes continue to transmit a health message every round, while the non-CH nodes aggregate the CH health messages every cycle and responds with a health message at the end of a cycle. 
%The \emph{health/acknowledge} messages allow the nodes to detect node failures, or if new nodes are within radio range. 
%A node is marked as faulty if it fails to transmit health message for 5 consecutive activity windows defined to be $T_{fail}$. 

%However, as the nodes activate RDC, transmission from other leaf nodes might not be received by all nodes due to the asynchronous sleep windows at each receiving node, leading to wrong conclusions about a nodes active state.
%Each of the above Discovery, Election and Correction phases completes in a round yielding only a single successful message transmission per phase as phase transitions happen at round boundaries.
Following the Discovery, Election and the Correction phases, all the nodes switch to the RDC mechanism and have different transmission intervals depending on their CH/non-CH status.
Furthermore, as the nodes are not synchronized, the sleep windows at each node might be different.
Thus, transmissions from a node may be missed by its neighbors, leading to false positives about a node's state.
%Missed messages could lead to false positives about a node's state, resulting in instability.

To overcome this, DeCoRIC employs a modified gossiping scheme, derived from~\cite{gossip}, to spread information about the node health.
Each node maintains a \emph{fail} counter (counting rounds) $T_{\mathrm{fail}}$ for every node in its neighbor list and it is incremented at every round.
Any node which receives a \emph{direct} message from a neighbor node resets the corresponding fail counter to zero and gossips about the health of the neighbor node by including its ID in its connectivity list. % transmitted during its health message. 
%When a node receives a direct message from its neighbor, it resets the corresponding fail counter (counting rounds) to zero. 
Meanwhile, if a node receives a \emph{gossip} message about a neighbor node (i.e., from the connectivity list of a received message), the fail counter corresponding to that node is reduced by half. %and is indicated in lines 5-10 of Algorithm~\ref{alg.cluster4}.
When the fail counter corresponding to a neighbor reaches $T_{\mathrm{fail}}$ at a node, the neighbor ID is marked faulty and removed from the connectivity list and, thus, its health message.
%However, the gossip is only re-initiated upon reception of a direct message and not the gossip message.
%This prevents perpetual gossiping of a failed node.

Assuming the hard fault model, once the fail counter reaches $2 \cdot T_{\mathrm{fail}}$, the corresponding node's ID is marked as failed and removed from its neighbor list. %the node as failed. 
The stable phase operation and the failure detection process is shown in Algorithm~\ref{alg.cluster4}.
The gossiping scheme helps to accommodate false triggers caused by missed packets or transient network conditions, at the expense of increased detection time for node failures. 

\begin{algorithm}[t!]
	\scalebox{0.90}{
	\begin{minipage}{0.98\columnwidth}
	\caption{Stable phase}
	\label{alg.cluster4}
		\begin{algorithmic}[1]
			\If {rcv()}	
			\State Msg = rcv().data
			\If{Msg.Degree \textgreater CH.Degree}
			\State	CH.ID = Msg.ID
			\State  Phase = ELECTION
			\EndIf
			\For {each item $i$ in Conn}
			\If {(Msg.Conn[i] $\&$ Conn[i])}
			\State fail[i]\,=\,0
			\ElsIf {(Msg.Conn[i] $\&$ !Conn[i])} 
			\State fail[i] = 0.5 $\cdot$ fail[i]
			\EndIf
			\EndFor
			
			\State Neighbor[Msg.ID], Conn[Msg.ID] = TRUE
			\EndIf
			
			\If{round}
			\For{each item $i$ in Conn}
				\If {fail[i] $\ge$ $T_{\mathrm{fail}}$}\,Conn[i] = FALSE
				\EndIf
				\If {fail[i] $\ge$ $2 \cdot T_{\mathrm{fail}}$}\,Neighbor[i] = FALSE
				\Else \,fail[i] = fail[i] + 1
				\EndIf
			\EndFor
			\If{node.ID == CH.ID} \,broadcast(Msg)
			\ElsIf {cycle} \,broadcast(Msg)
			\EndIf
			\EndIf
			
		\end{algorithmic}
	\end{minipage}
	}
\end{algorithm}

\paragraph*{Network Adaptation}
A failure of a node or addition of new nodes creates a change in the clustering of the network. The change ranges from a few clusters (new node addition or non-CH failure) to the entire network (CH failures) depending on the connectivity of the failed node to other clusters. 
Only nodes whose degree change as a result of node failure switch to the Election phase; unaffected nodes continue to operate in the Stable phase.
Finally, when a new node tries to integrate into an existing cluster (i.e., observing health messages), it joins into an existing cluster and could replace the current CH based on its degree that becomes apparent over the next cycle using the~\emph{New CH ID} field.

In our example network, when node 4 was deleted, it was observed that both nodes 12 and 13 go into the Election phase after the detection of the failure.
After the Election phase, node 12 declares itself as the new CH while node 13 operates as non-CH in the new cluster.
%Hence in a large network, only the affected cluster elements will switch their state, while unaffected nodes continue in stable operating state.

%The clusters formed remain stable and continue to operate with the elected CHs with no changes in Correction phase~\ref{phase3}.
%If there is a change in CH, the impacted nodes switch to Election phase again to form new clusters/CH.
%Hence, the algorithm strives to form stable clusters, reducing the number of unclustered nodes.
%Finally, when a new node tries to integrate into an existing cluster (i.e., it observes health messages instead of ping messages), it joins into an existing cluster and could replace the current CH based on its degree that becomes apparent over the next cycle. % the state switches occur over a minimum period of one cycle.
% There could also be a re-clustering triggered if there are network changes such as addition of a new node or removal of any node. 
% Once the network is stabilized, the CH and non-CH nodes transmission rates change to reduce power consumption. 
%CH nodes continue to transmit every round while non-CH nodes transmit every cycle.

% \iffalse
% \begin{definition}[]

% \end{definition}
% \fi

%\begin{equation}
%\nonumber 
%\end{equation}

\section{Analysis \& Evaluation}
\label{sec.experiments}

In this section, we present the evaluation of DeCoRIC using the Cooja Simulator from Contiki~\cite{Dunkels:2004}.
We implemented LEACH and BEEM protocols on Contiki for comparison with DeCoRIC. 
%All three protocols are evaluated with multiple experiments in the same environment.
We measure the average power consumption per node and the time for the first node death to compare the power efficiency of the protocols.
The protocol with the least power consumption and the longest time to death for the first node would have the most power-efficient operation, assuming they offer similar connectivity.
%We conduct multiple experiments to compare the Average power consumption, time for node death and energy consumption for connectivity across all the three protocols.}
%We implemented LEACH clustering protocol in Contiki for comparison with DeCoRIC.
%The simulation parameters used for our test setup are shown in Table~\ref{tab.simparams}.
%We measure the average power consumption of all clustering algorithms over a period of time.
%Further, as the nodes exhaust their energy, we observe the time for the death of first node in the network.
We also show the progression of nodes exhausting their energy over time to quantify the power distribution of the protocols among the nodes of the network.
This experiment also gives a measure of time during which the network stays intact and connected.
To further illustrate the connectivity, we reduce the range of nodes in the simulation to show the time taken and power expended by the protocols to achieve 100\% connectivity.
Additionally, our test scenarios analyze and evaluate the resilience of DeCoRIC by triggering faults in the network and quantifying the worst-case delay before the network stabilization post re-clustering.
%In this case, we quantify the time taken from the point of node failure to detection of failure.

\paragraph*{Cooja Simulator}
Contiki's Cooja Simulator allows development in native C language, which can then be directly deployed on a compatible hardware platform~\cite{Dunkels:2004}.
The software elements are cross-compiled to a target hardware, similar to an emulation flow.
This enables the evaluation to consider actual hardware constraints such as memory limitations (to fit the algorithm), network errors such as packet-loss and interference, and actual bit-level transmission at the cost of slower execution time. 
%If the memory footprint of the algorithm is large, it will not fit within the node's limited resources.
%There is a real-world network traffic simulation with packet-loss and interference, as nodes are emulated at hardware level. 
%This causes the simulation to be slow but gives a more precise estimation of parameters. The same firmware can directly be loaded into the physical nodes.
We use the Skymote~\cite{skymote} as the target hardware platform and employ the powertrace tool in Contiki to measure the power consumption of the devices for all our experiments. %~\cite{powertrace}
Skymote uses a Texas Instruments CC2420 transceiver that complies with the 2.4\,GHz IEEE\,802.15.4 6LoWPAN standard with a bit rate of 250\,kbps and a processor platform that supports sustained low-power mode.
%The hardware supports a sustained low-power operating mode for the CPU and features a Texas Instruments CC2420 transceiver allowing the use of the 2.4~GHz IEEE~802.15.4 6LoWPAN standard for communication at 250~kbps. 
%We also tested the Z1 mote~\cite{z1mote} platform for a set of nodes to validate DeCoRIC in a heterogeneous network. 
%We also tested the Z1 mote~\cite{z1mote} as a target hardware to verify the capability of the algorithm to work in a heterogeneous environment.
%Contiki uses Radio Duty Cycling (RDC) to keep the nodes in low power mode for longer time, achieving high power saving.

\paragraph*{LEACH and BEEM Implementation}
The original LEACH and BEEM implementations were done in MATLAB which abstracts away the low-level communication details (hardware radio model).
Also, the MATLAB implementations were inherently centralized since the simulation system has an overall view of the state of each node. 
Hence, we implemented LEACH and BEEM on Contiki based on the original protocol in MATLAB and the description in the papers~\cite{926982,6878886}~\footnote{For the first time in a decentralized system with hardware emulation.}.
At the lowest level, we use the Contiki radio model as a common platform for emulating LEACH, BEEM and DeCoRIC using the Cooja Simulator; the higher layers are the C implementations of the respective protocols.
All the protocol implementations in C are available as open-source for research use~\footnote{The source code is available at https://bitbucket.org/nitinshivaraman/clustering\_contiki.}.
%Due to a common platform, we were able to forgo the radio modeling described in~\cite{926982}.

%LEACH also performs clustering over a series of phases consisting of Cluster Head (CH) Election and Advertisement, Cluster formation with the elected cluster heads and Non-Cluster Head (non-CH) nodes and a steady phase with TDMA among the cluster nodes.
The radio of the non-CH nodes on both LEACH and BEEM implementations are turned off once the clustering is complete, except when they have to transmit messages to the CH.
Meanwhile, the radio of the CH is always kept on to receive messages from non-CH nodes of the cluster as described by the protocols.
We translate this TDMA behavior into Contiki using the RDC mechanism.
%We map this TDMA behavior using the same RDC mechanism supported by the Contiki radio model.

Although both protocols differ in their CH election process, they have a cyclic re-clustering mechanism after a period defined as an \emph{epoch}~\cite{926982,6878886}.
%Both protocols differ in their CH election process which has been implemented according to the description in the respective papers~\cite{926982,6878886}.
%LEACH and BEEM have a cyclic re-clustering mechanism after a period defined as an \emph{epoch}.
An optimal value of the epoch is paramount for energy efficiency on both protocols.
We conducted experiments to measure the power consumption by varying different epoch values. 
It was found that higher epochs yield a lower power consumption with CH nodes expending higher power, while shorter epochs lead to constant re-clustering and higher power expenditure from all nodes.

\begin{table}[t]
	\begin{minipage}[b]{1.0\linewidth}\centering
		\begin{center}
			\caption{Parameters used in our experimental setup for evaluating DeCoRIC against LEACH and BEEM.}
			\label{tab.simparams}
			\scalebox{0.9} {
				\begin{tabular}{@{}ll@{}}
					\toprule
					\textbf{Parameter}          & \textbf{Values used}   \\ \midrule
					Area (m\textasciicircum{}2) & 100x100                \\ 
					Number of nodes (N)             & \{50, 100, 200\}      \\ 
					Transmission Range (m)      & 50                     \\ 
					CDMA MAC Protocol           & CSMA-CA (CXMAC), TDMA        \\ 
					Radio Frequency (GHz)       & 2.4                    \\ 
					Topologies					& \{Random\}	 \\
					$maxBE$                     & 3 \\
					Packet rate (packets/node/round)     & 1 round                \\
					RDC rate (activations/s)    & 32                     \\ \bottomrule
				\end{tabular}
			}
		\end{center}
	\end{minipage}
\vspace{-5mm}
\end{table}

\paragraph*{Experimental setup}
All the experiments were performed on the Cooja simulator with different parameters. 
DeCoRIC, LEACH and BEEM are run for a group of 50, 100 and 200~nodes arranged in 100 random topologies in an area of 100~x~100~$\mathrm{m}^\mathrm{2}$.
The topologies are common for all the three protocols, with nodes placed at random locations within a given area using the random placement feature of the Cooja simulator.
%Both DeCoRIC as well as LEACH were tested on the same 100 random topologies generated through the simulator.
%The area was set to 100m x 100m for all experiments unless specified.
%Optimum transmission range of all nodes was set at 75m as specified in the standard~\cite{techreport}.
The packet transmission rate is 1 packet/round with a transmission range of 50~m for each node.
Round in DeCoRIC, is set based on Equation~\ref{eq:round} as 0.8, 1.1 and 2.2~seconds, respectively, for 50, 100 and 200~nodes.
The lower bound of 0.8~seconds is a restriction imposed by the simulator, below which transmission overlap was observed due to incomplete initialization, resulting in unintended collisions and data loss.
One cycle is configured as 6 rounds, $T_{\mathrm{fail}}^{\mathrm{CH}}$ and $T_{\mathrm{fail}}^{\mathrm{nCH}}$ for CH and non-CH nodes are computed as 6 and 36 rounds, respectively, using Equations~\ref{eq:round} and~\ref{eq:cycle}.
%In order to have a stable phase time similar \changes{among all the algorithms}, the epoch was chosen as 10 rounds. %to the Stable phase time of DeCoRIC
%Since LEACH has 4 phases, each taking up 1 round between phase transitions, there would be 6 rounds of Stable phase before the next phase of re-clustering begins.
To ensure that all three protocols achieve comparable stable state duration before the cyclic re-clustering process, the epoch was chosen to be 10 rounds.
%\changes{As DeCoRIC has transmissions at all nodes at the cycle boundary, the re-clustering interval for LEACH and BEEM was configuration was set to 6 rounds to facilitate equivalent conditions for experiments.}
%This is in congruence with the round configuration chosen for DeCoRIC where transmissions in all nodes occur at the cycle boundary which is configured to be 6 rounds as elaborated in Section~\ref{phase4}.
%Hence, this creates equivalent conditions for DeCoRIC and LEACH without any unfair advantage to either of the two algorithms.
%All the experimental results are plotted as box plots with a mean line to observe the variation across different topologies.
The simulation parameters used for our test setup are shown in Table~\ref{tab.simparams}.

\subsection{Power Consumption}
\begin{figure}[t!]
	\centering
	\includegraphics[width=1.0\columnwidth]{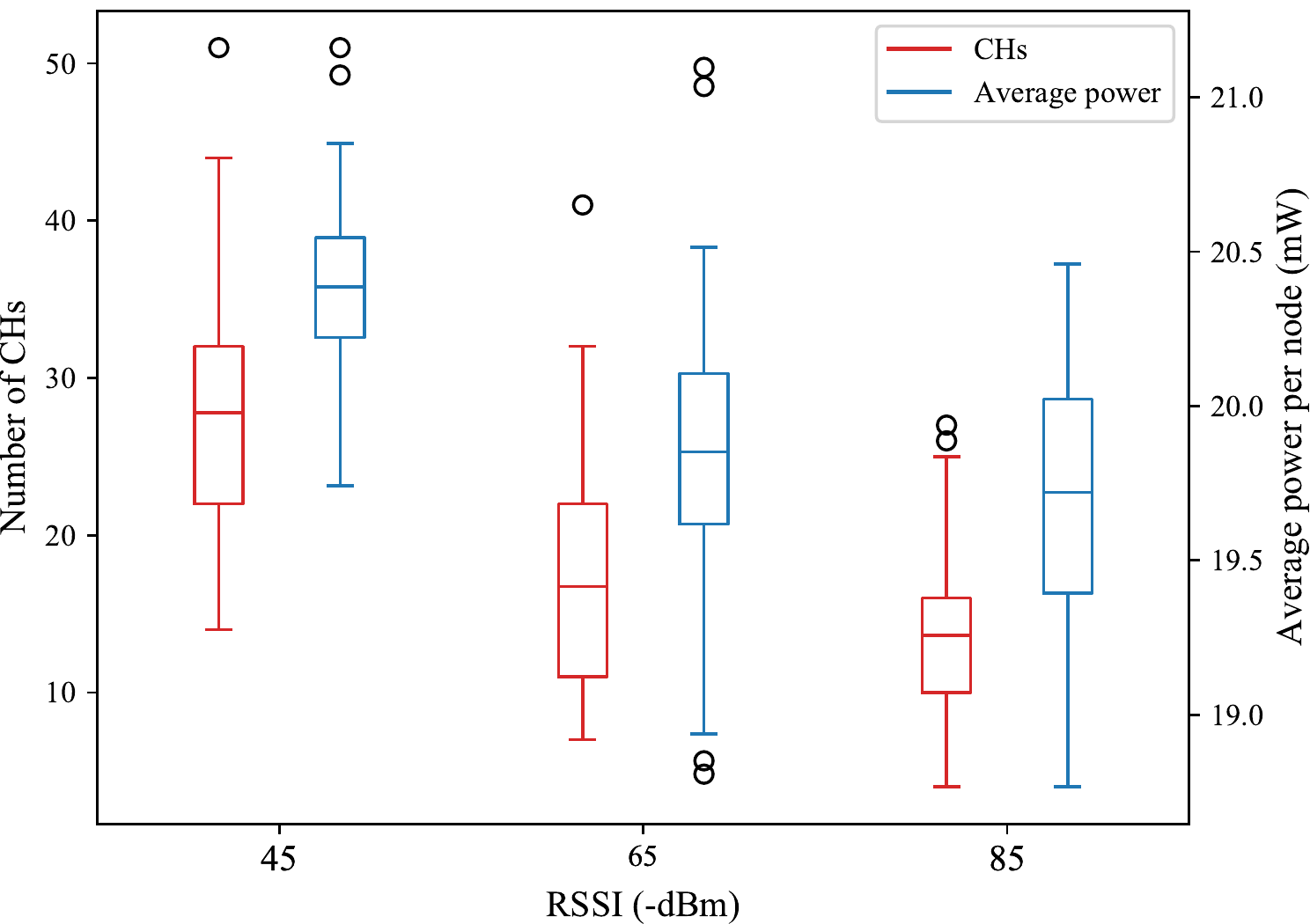}
	\caption{Number of CH nodes and average power consumed by the nodes running DeCoRIC over different RSSI thresholds to form clusters.} %The box plots show the variation in power consumption between inactive and active nodes in the system.} %Simulation performed with 50, 100 and 200 nodes with 1 packet every 10 seconds.}
	\label{fig.Energy}
\end{figure}

%We quantify the efficiency of DeCoRIC by evaluating the number of CH nodes, and the average power consumption of the network by varying configurable parameters in DeCoRIC. 
%The power consumption is averaged over 100 runs of random placement of nodes (i.e., 100 random topologies) \changes{for a time period of 1000 seconds.}
%Simulation is run  to measure the average power for 100 nodes in the network for the 100 random topologies.
Initially, we evaluate the change in average power consumption of the network and the resulting number of CH nodes by varying the RSSI value in DeCoRIC.
%a change in the number of nodes treated as cluster members or external neighbors as described earlier in Section~\ref{phase1}.
%Since the CH is primarily responsible for communication with other clusters and maintaining the sanctity of the cluster, a lower number of CHs translates to larger clusters.
%This leads to less messages transmitted externally which is reflected as power savings.
Figure~\ref{fig.Energy} shows the results of the experiment across different RSSI reception thresholds of -45\,dBm, -65\,dBm and -85\,dBm represented by the x-axis for 100 nodes. % in an area of 100$\times$100 m\textsuperscript{2}.
The y-axis on the left represents the number of CHs formed while the average power consumption per node in milliWatts (mW) is shown on the y-axis to the right.
%number of nodes in the topologies and remains same through the rest of the experiments.
The results show that in the case of -45\,dBm nodes (lower effective radio range), less than 30\% of nodes act as CH nodes on average with a worst-case of 43\% CH nodes. 
This is a result of many nodes being marked as external neighbors in this case due to their positions in the topology. 
%more than half the nodes could act as CH within clusters (excluding bridge-CHs), as many nodes are marked as external neighbors during the neighbor discovery phase. %reducing the overall power consumption of the network.

As the RSSI threshold increases, DeCoRIC marks more nodes as potential neighbors, with a mean and worst case of 13 and 25 CHs at -85\,dBm; a mean and worst case of 17 and 33 CHs at -65\,dBm.
In comparison, most clustering schemes (including LEACH and BEEM) predefine the number of CHs to be between 5--25\%~(see~\cite{6840926,Ari:2016,1045297,avril2014clustering}) of the total number of nodes with no consideration for the network structure, often resulting in disconnected clusters.
The outliers in the number of CHs for DeCoRIC  can be attributed to the randomness of the topologies since nodes that are farther than the radio range of the RSSI threshold become members of different clusters. 

Changing the RSSI results in a change of the cluster size, with higher RSSI leading to bigger clusters and lower RSSI leading to smaller ones.
Larger clusters expend higher energy on CHs while reducing the overall network power consumption; smaller clusters result in higher network power consumption with many CH nodes as seen in Figure~\ref{fig.Energy}.
The change in power is more pronounced as the number of nodes scales.
%In contrast, DeCoRIC dynamically determines the number of CHs through node activity and configures RSSI thresholds in a decentralized manner. Hence, it strives to achieve full network connectivity through its CH/bridge-CH combinations.
%Better clustering in DeCoRIC results in reduction of average power consumption across nodes in the network, as shown by the downward trend in the average power consumption plot in Figure~\ref{fig.Energy}.
Further, the outliers in the average power can be attributed to the fact that DeCoRIC employs bridge-CHs to facilitate connectivity in the network, which increases the average power consumption.
%We also observed that in topologies that have spatially separated regions of high node density, DeCoRIC creates more bridge-CHs to ensure connectivity between them, resulting in slightly higher average power consumption.

\begin{figure}[t!]
	\centering
	\includegraphics[width=1.0\columnwidth]{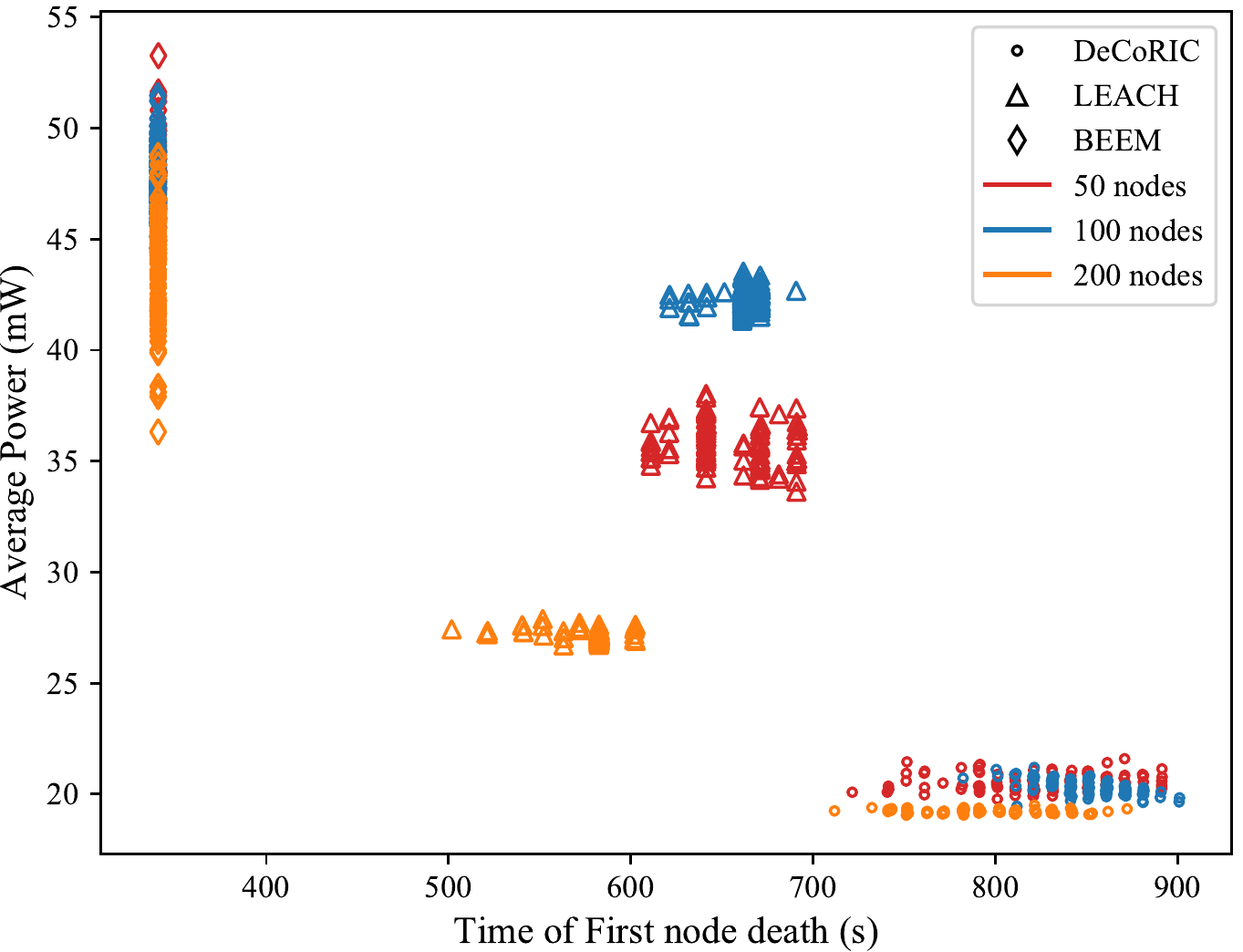}%{figures/scatter.pdf}%{figures/energy_comparo2.pdf}
	\caption{Average Power Consumption vs Time of first node death in the network for DeCoRIC compared against LEACH and BEEM.} %Simulation run for 5000 seconds with a packet rate of 1 packet per second for comparing against competing schemes.}
	\label{fig.EnergyComparo}
\end{figure}

%We also compare the average power consumption per node for DeCoRIC, LEACH and BEEM techniques. %and the power consumed during clustering phase 
%For a fair comparison, we have implemented LEACH protocol in a contiki environment.
%The experiment is setup using the parameters as described before in Table~\ref{tab.simparams}. 
The RSSI threshold of -65\,dBm is chosen for the rest of the experiments for DeCoRIC.
To demonstrate the power saving in DeCoRIC, first, we compare average power per node in milliWatts (mW) along with the time of their first node death (exhaust nodes' power completely) over a simulated duration of 1000 seconds for 50, 100 and 200 nodes in the network across LEACH, BEEM and DeCoRIC.
%The parameters are set based on the experimental setup described earlier comparing average power consumption in milliWatts (mW) in 1000 seconds over the number of nodes in the network.
Second, we record the time at which the nodes' death.
To identify residual energy in the Contiki framework, the power model in~\cite{riker2017iscc} was integrated with \emph{powertrace} with an initial battery capacity of \SI{6}{\milli Wh} (milli Watt Hour) to observe the energy drain of the nodes.
%Non-CH nodes select the closest CH in LEACH as there is no RSSI threshold setting.
%Both algorithms use the same underlying radio transceiver block (CC2420) and its radio model available in Contiki.
%We make use of Contiki's \emph{powertrace} to measure power consumption averaged over the number of nodes in the network.

%\begin{figure}[t!]
%	\centering
%	\includegraphics[width=1.0\columnwidth]{figures/box_time.pdf}
%	\caption{Time taken for the death of first node in DeCoRIC and LEACH.} %The box plots show the variation in power consumption between inactive and active nodes in the system.} %Simulation performed with with 50, 100 and 200 nodes with 1 packet every 10 seconds.}
%	\label{fig.Firstnodedeath}
%\end{figure}

The results of the first experiment are shown in Figure~\ref{fig.EnergyComparo}, where the x-axis and y-axis represent the time for the first node death in seconds and the average power consumption per node in mW respectively.
The number of nodes in the network is represented by different colors, while the protocols are represented by different shapes.
%We chose these 3 combinations since adding more combinations would impact the readability of the figure.
From the results, it is seen that our proposed method has the least power consumption per node, thereby prolonging the time for the first node death.
It offers a best-case of 70\% and 110\% improved power efficiency over LEACH and BEEM for 50 nodes. 
%Nitin:\changes{Similarly, the power efficiency is improved by x and y for 100 nodes and x and y for 200 nodes over LEACH and BEEM protocols.}
Similarly, the best-case improvement for the time of first node death is 42\% and 109\% over LEACH and BEEM for 200 nodes, respectively.
%Nitin:\changes{Likewise, an improvement of x and y for 50 and x and y for 100 nodes is seen.}
%Similarly, for 50 and 100 nodes in the network, it offers an improvement of x\% and y\% over LEACH and x\% and y\% over BEEM in the average power.
%It also offers 63\% and 81\% improvement in average power consumption of 50 and 100 nodes in a network respectively.

The energy savings in LEACH can be attributed to the proactive load distribution strategy with periodic re-clustering.
Hence, there is a longer time before the first node dies as the power distribution is balanced.
BEEM, on the other hand, adopts a strategy where CH nodes remain unchanged during re-clustering to retain connectivity, while non-CH nodes are retained in low-power mode during the periodic re-clustering.
Due to this strategy, the CH nodes exhaust their power rapidly due to prolonged radio on-time.
The periodic clustering in LEACH and BEEM results in higher power consumption for all the nodes.
%CH nodes and lower for non-CH nodes as the number of nodes increase. 
This is reflected in the plot where the total average power decreases as the number of nodes scales while the death of the first node happens faster.

%, where CHs are changed periodically expending considerable energy due to the re-clustering. 
%Furthermore, LEACH also requires considerably more messages as each node exchanges information about its residual energy to elect the next CH.
By contrast, DeCoRIC uses a reactive strategy, reducing the activity in the Stable phase and re-clustering only for node failures.
As there is no TDMA, all the nodes experience similar radio activity subject to the density of nodes in the network.
Similar to LEACH and BEEM, there is a decrease in power consumption and faster depletion of node power as the network scales.
Both BEEM and DeCoRIC retain the CH in order to ensure connectivity.
However, DeCoRIC balances the radio activity efficiently with RDC, re-clustering only after node failures are detected and strives to achieve maximum connectivity.
As seen from Figure~\ref{fig.EnergyComparo}, the power consumption and the active time of nodes are inversely related.

\begin{figure}[t!]
	\centering
	\includegraphics[width=1.0\columnwidth]{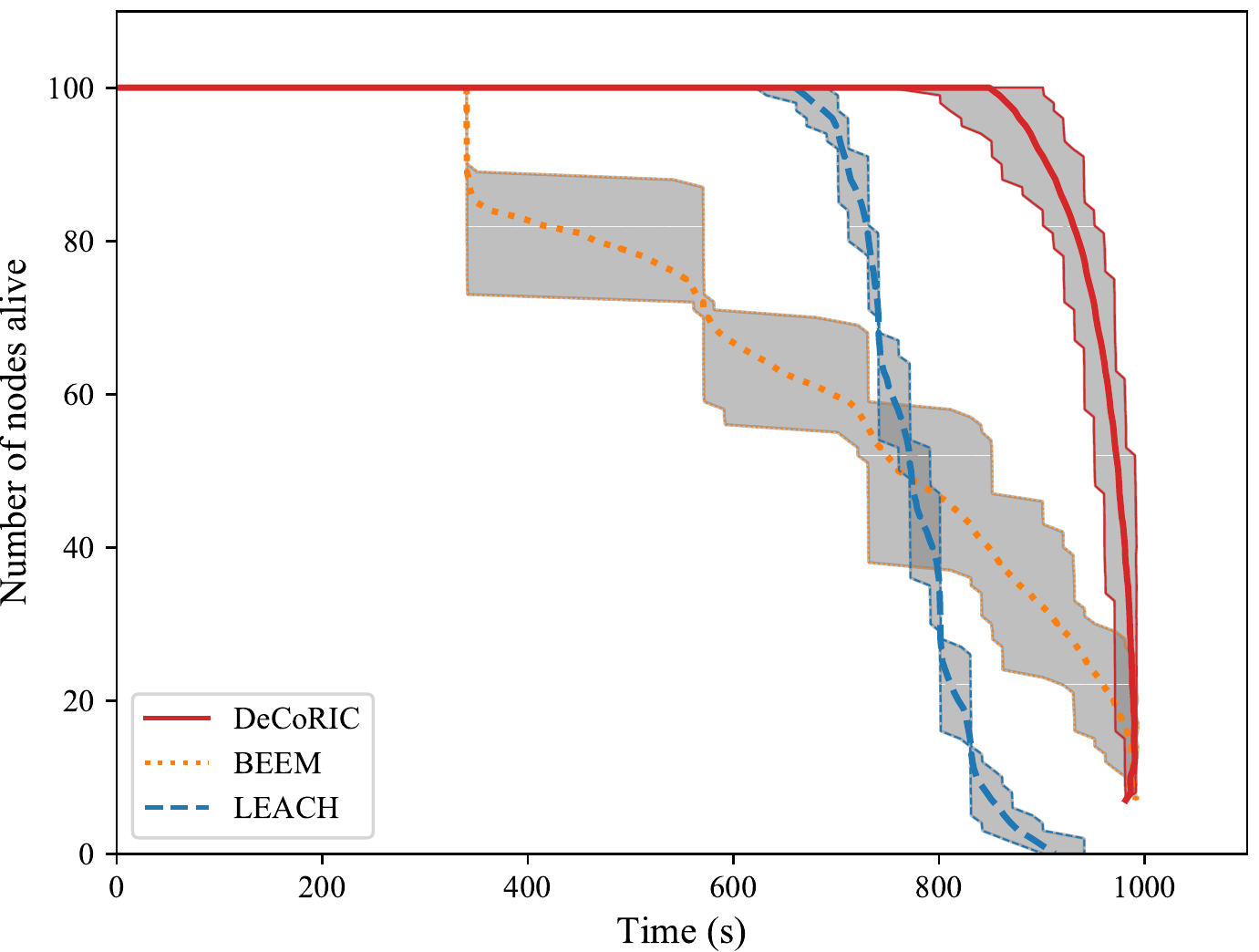}%{figures/box_power.pdf}%{figures/energy_comparo2.pdf}
	\caption{Battery drain of the network of DeCoRIC, LEACH and BEEM. The gray area indicates the variation between the minimum and maximum boundaries with the solid line representing the average.} %Simulation run for 5000 seconds with a packet rate of 1 packet per second for comparing against competing schemes.}
	\label{fig.batterydrain}
\end{figure}

%After the clustering process, there is a significant difference in the power consumption of CH nodes and non-CH nodes in LEACH and BEEM protocols.
%Relying on the failure detection mechanism that uses gossiping, DeCoRIC is able to employ RDC on both CH and non-CH nodes.
%As a result, the power consumption is dependent on the degree of a node, where the radio is kept on for incoming messages.
In order to maintain connectivity over a longer period, the power dissipation has to be managed efficiently among all the nodes.
%LEACH achieves this objective by distributing the CH status among all the nodes.
%BEEM aims to keep nodes alive for a longer time; the strategy focuses on CH nodes in areas of higher node density to retain the CH status to maintain connectivity. 
%However, this process also causes critical (bridge) nodes also to exhaust power in a faster rate.
%DeCoRIC balances power consumption through RDC among all nodes, maintaining the connectivity for a longer time.
To quantify the rate of power consumption and the time of connectivity, we show the time at which the nodes exhaust their energies in a simulation of 1000 seconds.
The time at which nodes die progressively is shown in Figure~\ref{fig.batterydrain}.
The y-axis of the plot represents the number of nodes in the network at the start of the simulation while the x-axis represents the time in seconds.

%\changes{LEACH has frequent re-clustering where CHs are rotated among different nodes balancing the load of CHs and distributing the power dissipation. -- is this repetitive?}
Energy exhaustion rate of nodes is higher in LEACH than BEEM and DeCoRIC, as most nodes would have expended similar energies.
%As BEEM has CH nodes retaining its status at every epoch, the energy is depleted rapidly causing the first node to die the earliest.
BEEM has certain nodes in a denser area that start consuming energy after the death of some CH nodes, leading to a longer lifetime for these nodes.
However, since nodes exhaust their energy at an early stage in BEEM, some key bridge nodes could exhaust energy quicker than the other nodes, leading to a disconnected network.
DeCoRIC manages power more efficiently using RDC, providing a longer time for the network to stay connected before the nodes exhaust their powers.
Since both DeCoRIC and BEEM aim to achieve connectivity, we see that the number of active nodes at the end of the experiment is similar for both algorithms.
%DeCoRIC also follows a similar pattern with LEACH initially but has a slower rate of nodes dying due to balanced use of radio and longer sleep times of non-CH nodes.

% and has a slower rate of death as CH nodes have a higher energy consumption than the non-CH nodes.}

\subsection{Connectivity}
We compare the clustering algorithms with respect to their ability to achieve connectivity among the nodes.
%In a network of N nodes, there are \textsuperscript{N}C\textsubscript{2} combinations of nodes.
%CHs and their cluster members form pairs for routing within a cluster.
%A CH forms a path with its neighbor CHs to route messages across clusters.
In order to test connectivity among nodes of the network, we reduce the transceiver range to 20m, as larger transmission range in a denser network enables all the nodes to be in communication range with each other.
This is a common strategy in dense networks for mitigating collisions, thereby reducing re-transmissions~\cite{4317902}.
Reduction in range enables reduction in transmission power which is the goal for most wireless sensors and IoT devices.

We measure the connectivity as the ratio of the number of connected nodes over the total $N$ nodes in the network.
Non-CH nodes of a cluster form a connected pair with its CH.
Similarly, neighboring CH nodes form a connected path among the clusters.
Combining such pairs, we get all the nodes that are connected in the network (where a routing path exists).
Depending on the topology, the maximum connectivity could vary from a single cluster covering a few nodes to multiple clusters covering all $N$ nodes of the network.
The former is a result when CHs are not in range of each other, forming independent clusters, and the latter is formed when all CHs are in range of one another to form a path among all $N$ nodes.

\begin{figure}[t!]
	\centering
	\includegraphics[width=1.0\columnwidth]{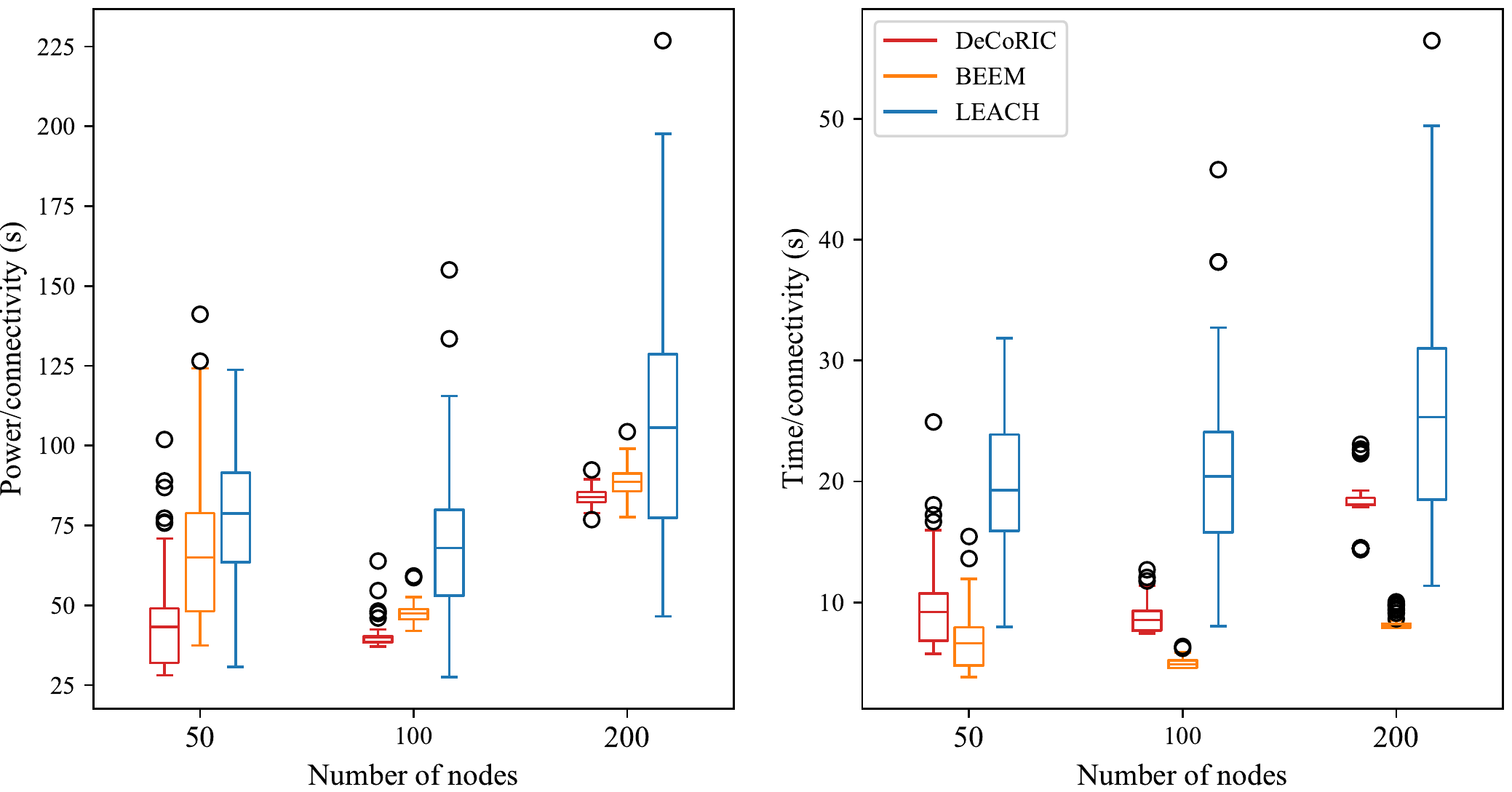}
	\caption{Power consumed and time taken for the network clustering normalized by connectivity among all the nodes in the network.} %The box plots show the variation in power consumption between inactive and active nodes in the system.} %Simulation performed with 50, 100 and 200 nodes with 1 packet every 10 seconds.}
	\label{fig.Connectivity}
\end{figure}

%\begin{figure}[t!]
%	\centering
%	\includegraphics[width=0.9\columnwidth]{figures/box_power_cluster.pdf}%{figures/energy_comparo2.pdf}
%	\caption{Initial Power Consumption during clustering in DeCoRIC and LEACH.}%Simulation run for 5000 seconds with a packet rate of 1 packet per second for comparing against competing schemes.}
%	\label{fig.Initpower}
%\end{figure}

While DeCoRIC and BEEM strive to achieve 100\% connectivity, DeCoRIC completes the clustering with less power and slightly longer time than BEEM.
LEACH consumes the least time and power for clustering but does not ensure connectivity.
Hence, in order to compare the performance of all the clustering schemes, we normalize both the power (power/connectivity) and time (time/connectivity) in the clustering phase by the connectivity achieved by the algorithms.

The results of the comparison are shown in Figure~\ref{fig.Connectivity}.
The x-axis represents the number of nodes in the network.
The y-axis of the left sub-plot represents the ratio of clustering power over connectivity while the y-axis of right sub-plot indicates the ratio of clustering time over connectivity.
As seen from the left sub-plot, DeCoRIC expends the least power to achieve 100\% connectivity, followed by BEEM and LEACH.
In contrast to the clustering power, DeCoRIC needs slightly longer to complete clustering compared to BEEM as shown in the right sub-plot.
The variations are attributed to the randomness of the topologies yielding different extents of connectivity.

As the number of nodes scales, the density of nodes increases, leading to better connectivity.
Although LEACH consumes the least power and time to complete clustering, the results normalized over connectivity show that LEACH would need much higher time to move towards 100\% connectivity.
%Additionally, since CHs are elected probabilistically in LEACH, the connectivity may change after an epoch due to the cyclic re-clustering.
Additionally, the 100 random topologies are representative of the changes in connectivity due to the change of CHs resulting from re-clustering changes.
BEEM also includes the cyclic re-clustering but retains the same CH to maintain connectivity.
This property of BEEM expends significant energy, retaining connectivity only while the CH nodes are alive. 
%As we saw in the previous experiment, BEEM has the fastest first node death, leading to faster loss in connectivity.

Similarly, the CHs are retained after the clustering is complete in DeCoRIC. 
However, over multiple epochs, the power consumption reduces significantly for DeCoRIC due to better radio management of the nodes.
The longer time of clustering in DeCoRIC is attributed to the Correction phase where the number of CH nodes is reduced while striving to attain 100\% connectivity.
The slightly longer clustering of DeCoRIC creates optimal clusters that sustain the connectivity for a longer time.
BEEM is faster as it does not consider the number of CH nodes active while achieving connectivity, thereby expending additional energy and leading to faster node deaths as seen in Figure~\ref{fig.batterydrain}.
Hence, overall connected time for BEEM is significantly lower than DeCoRIC, where DeCoRIC achieves over 2x longer connected time.

\subsection{Evaluation of Resilience}

%To evaluate resilience, we start with a stable network condition with 100 nodes that allows us to model the best and worst case node failures: a Bridge-CH node failure affecting multiple clusters, CH node failure, non-CH node failure and addition of a new node.
%For this setup, a round is configured as 1 second, one cycle as 6 rounds, $T_{\mathrm{fail}}^{\mathrm{CH}}$ and $T_{\mathrm{fail}}^{\mathrm{nCH}}$ for CH and non-CH nodes are computed as 6 and 36 rounds respectively.  
%In each case, we quantify the time taken by the network to detect the change, adapt and restart communication. The computed results are shown in Table~\ref{tab.resilience}.
%%The constant time taken to detect the failure of a node or the presence of a new node in the radio range is ensured by the activity window approach in DeCoRIC (see sec.~\ref{phase3}).
%%In case of CHs, the failure can be confirmed in 5 rounds (or 1 cycle) as the CH transmits the \emph{health} message every round and hence the nodes mark CH to have failed if the \emph{health} transmission is missed for 5 consecutive rounds. 

DeCoRIC does not assume any synchronization across the nodes and the round/cycle period in DeCoRIC aims to compensate for the lack of synchronization between the nodes, as explained in Section~\ref{sec.strategy}.
Since each round specifies a periodic set of actions (i.e., CH transmission, non-CH nodes receiving without any sequence order), the timing drift between nodes can be bounded to one round.
%RDC also influences successful message reception as the sleep time of a node may be aligned with the transmission window for the second node. 
%DeCoRIC employs gossiping to overcome this challenge, which allows a node's active state to be propagated by other nodes that receive a direct message.  
As explained in Section~\ref{phase3}, the failure window ($T_{\mathrm{fail}}$) covers the uncertainties caused due to the asynchronous RDC periods and transmission times, by defining $T_{\mathrm{fail}}$ as the least common multiple of the respective time periods. 
%Hence, DeCoRIC can declare a neighbor node to have failed if no message is received (either directly or via gossip) from this node for a time period of 2 $\cdot T_{\mathrm{fail}}$. 
%In an actual system, it is possible that a health message can be received anywhere within the $T_{\mathrm{fail}}$ window at a receiving node. 
%In the case of a node failure, no direct message is received within $T_{\mathrm{fail}}$ at receivers, and they stop gossiping about the active state of the node. 
%When a node receives a gossip, its fail counter is halved if the gossip is received within the $T_{\mathrm{fail}}$ window. 
%In the best case, a neighbor node does not receive any gossip about a failed node within the $T_{\mathrm{fail}}$ period and thus, confirms failure within twice this period following the gossip algorithm. 

If a transmission is not received at its neighbor and the neighbor receives a late gossip before the fail counter expires (at $T_{\mathrm{fail}}$), then the counter is halved as it waits to see if it was a transient fault. 
Thus, in the worst-case, there can be an additional half period (of $T_{\mathrm{fail}}$) that a neighbor waits before declaring the node to have failed. 
The fail period depends on the activity rate of the node; for CH nodes, their failure can be detected within a shorter window ($T_{\mathrm{fail}}^{\mathrm{CH}}$) compared to non-CH nodes, since CH nodes transmit more often than non-CH nodes.
These bounds are thus enforced by the protocol and are shown in Table~\ref{tab.resilience}.
Once a failure of CH or Bridge-CH node is detected, the nodes switch to the Election phase immediately and complete the recovery process over the next 2 rounds. 
In the case of a non-CH node failure, the recovery is immediate as there are no changes triggered in the cluster itself.

\begin{table}[t]
	\begin{minipage}[b]{1.0\linewidth}\centering
		\begin{center}
			\caption{Best and worst-case reaction time at each node with DeCoRIC in case of network changes.}\label{tab.resilience}
			\scalebox{0.9} {
				\begin{tabular}{@{}llll@{}}
					\toprule
					\multirow{2}{*}{\textbf{Network change}}  & \multicolumn{2}{@{}c@{}}{\textbf{Detection time}} & \multirow{2}{*}{\textbf{Recovery time}}    \\ \cmidrule{2-3}
					& Best-Case                  & Worst-Case                  	& \\ \midrule
					Fail: nCH node 			     	 		& 2 $\cdot T_{\mathrm{fail}}^{\mathrm{nCH}}$ 	 & 2.5 $\cdot T_{\mathrm{fail}}^{\mathrm{nCH}}$  	& immediate \\ 
					Fail: CH node  		    				& 2 $\cdot T_{\mathrm{fail}}^{\mathrm{CH}}$     & 2.5 $\cdot T_{\mathrm{fail}}^{\mathrm{CH}}$     & 2 rounds \\ 
					Fail: Bridge-CH node	       			& 2 $\cdot T_{\mathrm{fail}}^{\mathrm{CH}}$       & 2.5 $\cdot T_{\mathrm{fail}}^{\mathrm{CH}}$     & 2 rounds \\ 
					Add: New node							& 3 rounds   				 & 1 cycle      				& [0 or 3] rounds\\			
					\bottomrule                                                       
				\end{tabular}
			}
		\end{center}
	\end{minipage}
\vspace{-5mm}
\end{table}

When a new node integrates into the cluster, it can start following a CH within 3 rounds by listening to its broadcast messages. 
However, the new node can only determine its own degree over the next cycle when other non-CH nodes transmit.
If the new node has a higher degree than the current CH, it will transition as the CH by setting the \emph{New CH ID} field of the message frame, causing affiliated nodes to switch to the election phase to complete recovery. 
Otherwise, the recovery is completed immediately, and the new node integrates as a regular non-CH node. 
We verified the bounds stated in Table~\ref{tab.resilience} using our simulations, starting with a stable network condition and random topologies.

Meanwhile, LEACH and BEEM do not have a failure detection mechanism within the protocol.
Clustering operation repeats after every epoch, providing an upper bound for time to re-cluster as there is no failure detection mechanism.
This property results in a constant recovery time for any CH node independent of the topology changes in the network.
However, the worst-case recovery time could be longer depending on the configuration of an epoch.
%The time for recovery into stable phase after the re-clustering is shown in Table~\ref{tab.resilience_recovery}
%The time for recovery into Stable phase after the re-clustering for all combinations of the network topologies takes 3 rounds for LEACH while taking 2 rounds for DeCoRIC.
%\toshreejith{To correct if any error above.}
%In contrast to longer time to reach the Stable phase after failure, DeCoRIC consumes significantly lower power than LEACH while ensuring the network is resilient to changes.

\section{Conclusions and Future Work}
In this paper, we proposed a power-efficient decentralized clustering technique that can dynamically detect and adapt to node failures at runtime while ensuring connectivity among the nodes.
The protocol enables identification of nodes which enable connectivity and strives to create clusters that are connected. % while consuming minimal power in the process.
DeCoRIC provides a resilient and reliable communication framework in a network of any topological structure.
We show that the network can re-organize to form new clusters while maintaining connectivity even in case of critical CH failures, with a deterministic latency.
We implemented the state-of-the-art benchmark clustering protocol LEACH as well as BEEM, the protocol for connectivity, in the Contiki simulator for comparison with DeCoRIC.

We demonstrate that the number of CHs that are elected independently in DeCoRIC is similar to the number of CHs decided apriori in centralized schemes.
We also showed that DeCoRIC achieves up to 70\% better power efficiency and 42\% longer lifetime compared to LEACH while achieving up to 110\% better power efficiency and 109\% longer network lifetime in comparison to BEEM. %DeCoRIC clustering scheme offers up to 2$\times$ improvement in power consumption and minimal node lifetime over the state-of-the-art decentralized scheme.
Connectivity is achieved among nodes even in sparse networks using less power by accepting a slightly longer time for the clustering phase compared to BEEM. % with up to 49\% more efficiently

In the future, we aim to further enhance the failure detection model in DeCoRIC and incorporate a time-synchronization framework which can aid in further improving energy consumption.
We also aim to improve the initial clustering power consumption and minimize the detection time based on a more reliable physical layer protocol.
Furthermore, we plan to deploy this technique on a hardware testbed as required in safety-critical IoT applications.

%\vspace{-0.5mm}


%\section*{Acknowledgment}
%This work was financially supported in part by the XXXXXXX under its XXXXXXXX programme.
%This work was financially supported in part by the Singapore National Research Foundation under its Campus for Research Excellence And Technological Enterprise (CREATE) programme.




%\bibliographystyle{IEEEtran}
%\bibliography{IEEEabrv,fpl14_refs}
% \end{thebibliography}

%{\footnotesize
%\fontsize{8.6}{9.0}\selectfont
\bibliographystyle{IEEEtran}
\bibliography{all}
%}
%\input{bio}

\end{document}